\documentclass[twocolumn]{aastex61}

\usepackage{multirow}

\received{\today}
\revised{\today}
\accepted{\today}
\submitjournal{ApJ}

\shortauthors{Ertel et al.}

\begin{document}

\title{The HOSTS survey -- Exozodiacal dust measurements for 30 stars}

\correspondingauthor{Steve Ertel}
\email{sertel@email.arizona.edu}

\author{Ertel,~S.}
\affiliation{Steward Observatory, Department of Astronomy, University of Arizona, 993 N. Cherry Ave, Tucson, AZ, 85721,
USA}

\author{Defr\`ere,~D.}
\affiliation{Space sciences, Technologies \& Astrophysics Research (STAR) Institute, University of Li\`ege, Li\`ege,
Belgium}

\author{Hinz,~P.}
\affiliation{Steward Observatory, Department of Astronomy, University of Arizona, 993 N. Cherry Ave, Tucson, AZ, 85721,
USA}

\author{Mennesson,~B.}
\affiliation{Jet Propulsion Laboratory, California Institute of Technology, 4800 Oak Grove Dr., Pasadena, CA 91109,
USA}

\author{Kennedy,~G.~M.}
\affiliation{Department of Physics, University of Warwick, Gibbet Hill Road, Coventry CV4 7AL, UK}
\affiliation{Institute of Astronomy, University of Cambridge, Madingley Road, Cambridge CB3 0HA, UK}

\author{Danchi,~W.~C.}
\affiliation{NASA Goddard Space Flight Center, Exoplanets \& Stellar Astrophysics Laboratory, Code 667, Greenbelt, MD
20771, USA}

\author{Gelino,~C.}
\affiliation{Jet Propulsion Laboratory, California Institute of Technology, 4800 Oak Grove Dr., Pasadena, CA 91109,
USA}

\author{Hill,~J.~M.}
\affiliation{Large Binocular Telescope Observatory, University of Arizona, 933 N. Cherry Avenue, Tucson, AZ 85721, USA}

\author{Hoffmann,~W.~F.}
\affiliation{Steward Observatory, Department of Astronomy, University of Arizona, 993 N. Cherry Ave, Tucson, AZ, 85721,
USA}

\author{Rieke,~G.}
\affiliation{Steward Observatory, Department of Astronomy, University of Arizona, 993 N. Cherry Ave, Tucson, AZ, 85721,
USA}

\author{Shannon,~A.}
\affiliation{Department of Astronomy and Astrophysics, The Pennsylvania State University, State College, PA 16801, USA}
\affiliation{Center for Exoplanets and Habitable Worlds, The Pennsylvania State University, State College, PA 16802,
USA}

\author{Spalding,~E.}
\affiliation{Steward Observatory, Department of Astronomy, University of Arizona, 993 N. Cherry Ave, Tucson, AZ, 85721,
USA}

\author{Stone,~J.~M.}
\altaffiliation{Hubble fellow.}
\affiliation{Steward Observatory, Department of Astronomy, University of Arizona, 993 N. Cherry Ave, Tucson, AZ, 85721,
USA}

\author{Vaz,~A.}
\affiliation{Steward Observatory, Department of Astronomy, University of Arizona, 993 N. Cherry Ave, Tucson, AZ, 85721,
USA}

\author{Weinberger,~A.~J.}
\affiliation{Department of Terrestrial Magnetism, Carnegie Institution of Washington, 5241 Broad Branch Road NW,
Washington, DC, 20015, USA}

\author{Willems,~P.}
\affiliation{Jet Propulsion Laboratory, California Institute of Technology, 4800 Oak Grove Dr., Pasadena, CA 91109,
USA}

\author{Absil,~O.}
\affiliation{Space sciences, Technologies \& Astrophysics Research (STAR) Institute, University of Li\`ege, Li\`ege,
Belgium}

\author{Arbo,~P.}
\affiliation{Steward Observatory, Department of Astronomy, University of Arizona, 993 N. Cherry Ave, Tucson, AZ, 85721,
USA}

\author{Bailey,~V.~P.}
\affiliation{Jet Propulsion Laboratory, California Institute of Technology, 4800 Oak Grove Dr., Pasadena, CA 91109,
USA}

\author{Beichman,~C.}
\affiliation{NASA Exoplanet Science Institute, MS 100-22, California Institute of Technology, Pasadena, CA 91125, USA}

\author{Bryden,~G.}
\affiliation{Jet Propulsion Laboratory, California Institute of Technology, 4800 Oak Grove Dr., Pasadena, CA 91109,
USA}

\author{Downey,~E.~C.}
\affiliation{Steward Observatory, Department of Astronomy, University of Arizona, 993 N. Cherry Ave, Tucson, AZ, 85721,
USA}

\author{Durney,~O.}
\affiliation{Steward Observatory, Department of Astronomy, University of Arizona, 993 N. Cherry Ave, Tucson, AZ, 85721,
USA}

\author{Esposito,~S.}
\affiliation{INAF-Osservatorio Astrofisico di Arcetri, Largo E. Fermi 5, I-50125 Firenze, Italy}

\author{Gaspar,~A.}
\affiliation{Steward Observatory, Department of Astronomy, University of Arizona, 993 N. Cherry Ave, Tucson, AZ, 85721,
USA}

\author{Grenz,~P.}
\affiliation{Steward Observatory, Department of Astronomy, University of Arizona, 993 N. Cherry Ave, Tucson, AZ, 85721,
USA}

\author{Haniff,~C.~A.}
\affiliation{Cavendish Laboratory, University of Cambridge, JJ Thomson Avenue, Cambridge CB3 0HE, UK}

\author{Leisenring,~J.~M.}
\affiliation{Steward Observatory, Department of Astronomy, University of Arizona, 993 N. Cherry Ave, Tucson, AZ, 85721,
USA}

\author{Marion,~L.}
\affiliation{Space sciences, Technologies \& Astrophysics Research (STAR) Institute, University of Li\`ege, Li\`ege,
Belgium}

\author{McMahon,~T.~J.}
\affiliation{Steward Observatory, Department of Astronomy, University of Arizona, 993 N. Cherry Ave, Tucson, AZ, 85721,
USA}

\author{Millan-Gabet,~R.}
\affiliation{NASA Exoplanet Science Institute, MS 100-22, California Institute of Technology, Pasadena, CA 91125, USA}

\author{Montoya,~M.}
\affiliation{Steward Observatory, Department of Astronomy, University of Arizona, 993 N. Cherry Ave, Tucson, AZ, 85721,
USA}

\author{Morzinski,~K.~M.}
\affiliation{Steward Observatory, Department of Astronomy, University of Arizona, 993 N. Cherry Ave, Tucson, AZ, 85721,
USA}

\author{Pinna,~E.}
\affiliation{INAF-Osservatorio Astrofisico di Arcetri, Largo E. Fermi 5, I-50125 Firenze, Italy}

\author{Power,~J.}
\affiliation{Large Binocular Telescope Observatory, University of Arizona, 933 N. Cherry Avenue, Tucson, AZ 85721, USA}

\author{Puglisi,~A.}
\affiliation{INAF-Osservatorio Astrofisico di Arcetri, Largo E. Fermi 5, I-50125 Firenze, Italy}

\author{Roberge,~A.}
\affiliation{NASA Goddard Space Flight Center, Exoplanets \& Stellar Astrophysics Laboratory, Code 667, Greenbelt, MD
20771, USA}

\author{Serabyn,~E.}
\affiliation{Jet Propulsion Laboratory, California Institute of Technology, 4800 Oak Grove Dr., Pasadena, CA 91109,
USA}

\author{Skemer,~A.~J.}
\affiliation{Astronomy Department, University of California Santa Cruz, 1156 High Street, Santa Cruz, CA 95064, USA}

\author{Stapelfeldt,~K.}
\affiliation{Jet Propulsion Laboratory, California Institute of Technology, 4800 Oak Grove Dr., Pasadena, CA 91109,
USA}

\author{Su,~K.~Y.~L.}
\affiliation{Steward Observatory, Department of Astronomy, University of Arizona, 993 N. Cherry Ave, Tucson, AZ, 85721,
USA}

\author{Vaitheeswaran,~V.}
\affiliation{Steward Observatory, Department of Astronomy, University of Arizona, 993 N. Cherry Ave, Tucson, AZ, 85721,
USA}

\author{Wyatt,~M.~C.}
\affiliation{Institute of Astronomy, University of Cambridge, Madingley Road, Cambridge CB3 0HA, UK}




\begin{abstract}
The HOSTS (Hunt for Observable Signatures of Terrestrial Systems) survey searches for dust near the habitable zones
(HZs) around nearby, bright main sequence stars. We use nulling interferometry in N band to suppress the bright stellar
light and to probe for low levels of HZ dust around the 30 stars observed so far. Our overall detection rate is 18\%,
including four new detections, among which are the first three around Sun-like stars and the first two around stars
without any previously known circumstellar dust. The inferred occurrence rates are comparable for early type and
Sun-like stars, but decrease from $60^{+16}_{-21}$\% for stars with previously detected cold dust to $8^{+10}_{-3}$\%
for stars without such excess, confirming earlier results at higher sensitivity. For completed observations on
individual stars, our sensitivity is five to ten times better than previous results. Assuming a lognormal excess
luminosity function, we put upper limits on the median HZ dust level of 13\,zodis (95\% confidence) for a sample of
stars without cold dust and of 26\,zodis when focussing on Sun-like stars without cold dust. However, our data suggest
that a more complex luminosity function may be more appropriate. For stars without detectable LBTI excess, our upper
limits are almost reduced by a factor of two, demonstrating the strength of LBTI target vetting for future exo-Earth
imaging missions. Our statistics are so far limited and extending the survey is critical to inform the design of future
exo-Earth imaging surveys.
\end{abstract}




\section{Introduction} \label{sect_intro}
Exozodiacal dust -- exozodi for short -- is warm and hot dust (temperatures between few 100\,K and $\sim$2000\,K)
around main sequence stars. In analogy to zodiacal dust in the Solar system, the term refers to dust near the habitable
zone (HZ) of the host star, and closer in. It is produced through asteroid collisions \citep{der02} and comet
evaporation \citep{nes10}, and is redistributed under the influence of additional collisions, stellar radiation, wind,
and magnetic fields, as well as through interaction with any nearby planets (e.g., \citealt{wya05, sta08, bro09, rei11,
ert12b, vanlie14, ken15b, rie16}). Thus, studying the dust gives insight into the architecture and dynamics of
planetary systems in their inner regions, including the HZ.

At the same time, the potential presence of large amounts of HZ dust around nearby stars is a dominant source of
uncertainty for planning future exo-Earth imaging missions \citep{rob12}. The typical amount of dust present determines
the size of the primary mirror(s) needed to detect exo-Earths with a coronagraph or starshade in the visible (e.g.,
\citealt{sta15, sta16}), or a mid-infrared nulling interferometer \citep{def10}. Characterizing the occurrence rate of
the dust and its potential correlation with more accessible properties of the systems such as stellar spectral type,
age, or the presence of massive Kuiper belt analogs is vital for the design and target selection of such missions, and
thus for their success.

Because of its high temperature compared to colder, Kuiper belt-like debris disks, exozodiacal dust emits predominantly
in the near- and mid-infrared (nIR and mIR) where aside from a few exceptional cases it is outshone by the host star.
The ability of photometry and low resolution spectroscopy to disentangle disk and stellar emission are limited by
uncertainties from calibration and the prediction of the stellar photospheric flux, such that their typical sensitivity
limits are of the order of few percent of the stellar flux \citep{bei06}. Detecting scattered light from the dust
requires extreme contrast very close to the star. Targeted coronagraphic observations with the Wide-Field Infrared
Survey Telescope (WFIRST, \citealt{kri16}) may be able to image a few systems, but potential targets need to be
identified first. Given the small angular scales involved (1\,AU at 10\,pc corresponds to 0.1$''$), only infrared
interferometry currently provides the angular resolution and contrast needed to spatially disentangle the dust emission
from the stellar emission, enabling the detection of faint excesses. Optical long baseline interferometry in the nIR
has been very successful in detecting and characterizing hot dust very close to nearby main sequence stars
\citep{abs06, abs13, def12, ert14b, ert16, mar17, nun17}, but its connection to HZ dust is still unclear. On the other
hand, the emission of warm, HZ dust with a temperature of $\sim$300\,K peaks in the mIR, where nulling interferometry
is currently the most sensitive method to detect it. With this technique, the light from the central star is brought to
destructive interference, producing a central dark fringe, while spatially resolved emission is transmitted.

This method was used for a first dedicated exozodi survey by the Keck Interferometer Nuller (KIN, \citealt{mil11,
ser12, men13}). While critical constraints on the occurrence of bright exozodiacal dust were derived, the detection of
dust levels comparable to the Solar system was out of reach and the results could not sufficiently inform the design of
future exo-Earth imaging missions. To go beyond these results, we have developed the Large Binocular Telescope
Interferometer (LBTI, \citealt{hin16}) and its mIR nulling mode. We are carrying out the HOSTS (Hunt for Observable
Signatures of Terrestrial Systems, \citealt{dan14}) survey. In this paper we present the statistical results from the
30 individual stars observed so far. We provide the strongest constraints on the incidence rate and typical brightness
of HZ dust. They are particularly timely, because they provide valuable input for NASA's 2020 decadal survey, during
which mission concepts potentially capable of exo-Earth imaging will be evaluated (HabEx, \citealt{men16b}; LUVOIR,
\citealt{cro16}).

We discuss the sample of stars included in this paper in Sect.~\ref{sect_sample}. Our instrument and observing strategy
are described in Sect.~\ref{sect_inst_det}. The data reduction and detection method are described in
Sect.~\ref{sect_reduction}. Our results are presented in Sect.~\ref{sect_results} and discussed in
Sect.~\ref{sect_discussion}. Our conclusions are presented in Sect.~\ref{sect_conc}.

\begin{deluxetable*}{ccccccccccc}[]
\tablecaption{Observed sample as of June 2017.\label{tab_targets}}
\tablecolumns{12}
\tabletypesize{\footnotesize}
\tablehead{
\colhead{HD} & Name & \# SCI$^{~a}$ & Sp. Type & $V$ & $K$ & $N'^{~b}$ & $d$ & EEID$^{~c}$ & fIR/nIR & Excess\\[-5pt]
number & & & & (mag) & (mag) & (Jy) & (pc) & (mas) & excess & references
}
\startdata
\multicolumn{11}{l}{\textbf{Sensitivity driven sample}$^{~d}$\textbf{:}}\\
\hline
  33111   & $\beta$\,Eri   & 2  & A3\,IV      & 2.782  & 2.38  & 3.7  & 27.4  & 248  & N/N      & 1,2,3  \\
  81937   & 23\,UMa        & 3  & F0\,IV      & 3.644  & 2.73  & 2.6  & 23.8  & 168  & N/--     & 4      \\
  95418   & $\beta$\,UMa   & 4  & A1\,IV      & 2.341  & 2.38  & 4.2  & 24.5  & 316  & Y/N      & 5,6    \\
  97603   & $\delta$\,Leo  & 4  & A5\,IV      & 2.549  & 2.26  & 3.9  & 17.9  & 278  & N/N      & 1,2,6  \\
  103287  & $\gamma$\,UMa  & 4  & A0\,V       & 2.418  & 2.43  & 3.7  & 25.5  & 308  & N/--     & 1,2,5  \\
  106591  & $\delta$\,UMa  & 4  & A2\,V       & 3.295  & 3.10  & 2.0  & 24.7  & 199  & N/N      & 1,2,6  \\
  108767  & $\delta$\,Crv  & 2  & A0\,IV      & 2.953  & 3.05  & 2.3  & 26.6  & 251  & N/Y      & 1,2,3  \\
  128167  & $\sigma$\,Boo  & 3  & F4\,V       & 4.467  & 3.47  & 1.4  & 15.8  & 117  & Y/N$^{~e}$ & 1,6    \\
  129502  & $\mu$\,Vir     & 3  & F2\,V       & 3.865  & 2.89  & 2.6  & 18.3  & 151  & N/N      & 1,2    \\
  172167  & $\alpha$\,Lyr  & 2  & A0V         & 0.074  & 0.01  & 38.6 & 7.68  & 916  & Y/Y      & 5,7    \\
  187642  & $\alpha$\,Aql  & 2  & A7\,V       & 0.866  & 0.22  & 21.6 & 5.13  & 570  & N/Y      & 1,2,6,8 \\
  203280  & $\alpha$\,Cep  & 1  & A8\,V       & 2.456  & 1.85  & 7.0  & 15.0  & 294  & N/Y      & 1,2,6,8 \\
\hline
\multicolumn{11}{l}{\textbf{Sun like stars sample}$^{~d}$\textbf{:}}\\
\hline
  10476   & 107\,Psc       & 3  & K1\,V       & 5.235  & 3.29  & 2.0  & 7.53  & 90   & N/N      & 1,6,9,10 \\
  16160   & GJ\,105\,A     & 1  & K3\,V       & 5.815  & 3.45  & 1.5  & 7.18  & 73   & N/--     & 1,9,10  \\
  30652   & 1\,Ori         & 2  & F6\,V       & 3.183  & 2.08  & 4.8  & 8.07  & 205  & N/N      & 1,6,9,10 \\
  34411   & $\lambda$\,Aur & 2  & G1\,V       & 4.684  & 3.27  & 1.8  & 12.6  & 105  & N/--     & 10,11   \\
  48737   & $\xi$\,Gem     & 3  & F5\,IV-V    & 3.336  & 2.13  & 4.3  & 18.0  & 196  & --/N     & 6      \\
  88230   & GJ\,380        & 2  & K8\,V       & 6.598  & 3.21  & 1.9  & 4.87  & 65   & N/--$^{~f}$ & 12     \\
  89449   & 40\,Leo        & 2  & F6\,IV-V    & 4.777  & 3.65  & 1.1  & 21.4  & 98   & N/--     & 1,4    \\
  120136  & $\tau$\,Boo    & 2  & F6\,IV      & 4.480  & 3.36  & 1.7  & 15.6  & 114  & N/N      & 3,10,11 \\
  126660  & $\theta$\,Boo  & 3  & F7\,V       & 4.040  & 2.81  & 3.1  & 14.5  & 147  & N/--     & 1,9,10 \\
  141004  & $\lambda$\,Ser & 2  & G0\,IV-V    & 4.413  & 2.98  & 2.4  & 12.1  & 121  & N/N      & 1,6,9,13 \\
  142373  & $\chi$\,Her    & 3  & G0\,V       & 4.605  & 3.12  & 2.0  & 15.9  & 111  & N/N      & 1,4,6,9 \\
  142860  & $\gamma$\,Ser  & 4  & F6\,IV      & 3.828  & 2.63  & 2.9  & 11.3  & 151  & N/N      & 1,6,9,11 \\
  173667  & 110\,Her       & 2  & F6\,V       & 4.202  & 3.03  & 2.2  & 19.2  & 131  & Y/Y      & 6,12   \\
  185144  & $\sigma$\,Dra  & 2  & G9\,V       & 4.664  & 2.83  & 2.7  & 5.76  & 113  & N/N      & 6,10,11 \\
  215648  & $\xi$\,Peg\,A  & 3  & F6\,V       & 4.203  & 2.90  & 2.2  & 16.3  & 132  & N/N      & 1,4,9  \\
\hline
\multicolumn{11}{l}{\textbf{Commissioning targets:}}\\
\hline
  22049   & $\epsilon$\,Eri & 2 & K2\,V       & 3.721  & 1.66  & 7.4  & 3.22  & 172  & Y/N      & 7,14   \\
  102647  & $\beta$\,Leo   & 2  & A3\,V       & 2.121  & 1.92  & 6.9  & 11.0  & 336  & Y/Y      & 5,7    \\
  109085  & $\eta$\,Crv    & 3  & F2\,V       & 4.302  & 3.54  & 1.8  & 18.3  & 125  & Y/N      & 7,15   \\
\enddata
\tablecomments{Magnitudes are given in the Vega system.\\
$^a$~Number of calibrated science pointings obtained.
$^b$~Predicted flux in NOMIC $N'$~filter.
$^c$~Earth Equivalent Insolation Distance (Sect.~\ref{sect_opt_ap}).
$^d$~Sect.~\ref{sect_sample}.
$^e$~Mis-classified by \citet{gas13} as no excess.
$^f$~Cold excess \citep{eir13} likely background contamination \citep{gas14}. \\
References are:
Spectral type: SIMBAD;
$V$~magnitude: \citet{kha07};
$K$~magnitude: \citet{gez93} and the Lausanne photometric data base (http://obswww.unige.ch/gcpd/gcpd.html);
$N$~band flux and EEID: \citet{wei15};
Distance: \citet{vanle07};
Excess: (1)~\citet{gas13}, (2)~\citet{thu14}, (3)~\citet{ert14b}, (4)~\citet{bei06}, (5)~\citet{su06},
(6)~\citet{abs13}, (7)~\citet{abs06} (8)~\citet{rie05}, (9)~\citet{mon16}, (10)~\citet{tri08}, (11)~\citet{law09},
(12)~\citet{eir13}, (13)~\citet{koe10}, (14)~\citet{aum85}, (15)~\citet{aum88}.}
\end{deluxetable*}

\section{Sample}
\label{sect_sample}

The full target list of the HOSTS survey has been discussed in detail by \citet{wei15}. In short, it consists of
nearby, bright main sequence stars ($>1$\,Jy in $N$~band) without known close ($<1.5''$) binary companions. The sample
is separated into early type stars (spectral types~A to~F5) for which our observations are most sensitive and Sun-like
stars (spectral types F6 to K8) which are preferred targets for future exo-Earth imaging missions. The combination of
the two groups allows us to probe the incidence rate of exozodiacal dust across a large range of stellar masses,
thereby providing access to the physical processes at play for its production and evolution. In order to provide the
strongest intermediate results at any time, we maintain a balanced sample between early type and Sun-like stars during
the observations. Table~\ref{tab_targets} lists the basic, relevant properties of the targets observed so far. For a
fraction of the stars, the observations have been completed (three or four calibrated science pointings obtained). For
the other stars, more data will be obtained, so that the final null measurements are expected to differ within the
uncertainties from the values presented here and the final uncertainties will be smaller than the ones in the present
work.

Calibrators were selected following \citet{men14} using the catalogs of \citet{bor02} and \citet{mer05}, supplemented
by stars from the JSDC catalog and the SearchCal tool (both \citealt{che16}) where necessary. A minimum of three (in
most cases four) different stars were used to calibrate the observations of a single science target
(Sect.~\ref{sect_obs_strategy}) to minimize systematics due to imperfect knowledge of the calibrators (uncertain
diameters, potential multiplicity or faint circumstellar emission).

\section{Observations}
\label{sect_inst_det}

LBTI observations -- including high contrast direct imaging and integral field spectroscopy -- are scheduled
dynamically in queue mode to match observing conditions and project requirements. Most data presented in this work were
obtained between Sep.~2016 and May~2017 (LBT observing semesters 2016B and 2017A) as part of the HOSTS survey. In
addition, three stars were observed during LBTI commissioning: $\eta$\,Crv  (Feb.~2014, \citealt{def15}),
$\epsilon$\,Eri (Nov.~2014), and $\beta$\,Leo (Feb.~2015, \citealt{def16}; Hinz et al., in prep.). A brief log of the
observations is presented in Table~\ref{tab_obslog}. All raw and calibrated HOSTS data will be available to the public
one year after the observation date through the LBTI Archive (http://lbti.ipac.caltech.edu/).

\subsection{Instrument description}
The HOSTS observations are carried out using the LBTI \citep{hin16} at the Large Binocular Telescope (LBT). The
instrument combines the light from the two 8.4\,m apertures separated by 14.4\,m (center to center) on a common mount.
The two wavefronts are stabilized by adaptive secondary mirrors using two independent, closed loop adaptive optics
sub-systems (one for each aperture) with one pyramid wavefront sensor each, operating in the $R$~to $I$~band range
\citep{bai14}. The infrared light then enters the cryogenically cooled beam combiner. Active optical path delay (OPD)
and tip-tilt correction are performed using a closed-loop subsystem with a fringe tracker operating in K band. Active
vibration correction can be performed in the phase loop using telemetry from the Optical Vibration Measurement System
(OVMS, \citealt{boe16}) on the LBT. The mIR light (filter $N'$ for the observations presented in this work,
$\lambda_{\rm c} = 11.11\,\mu$m, $\Delta\lambda = 2.60\,\mu$m) is then combined in the pupil plane and re-imaged on the
Nulling Optimized Mid Infrared Camera (NOMIC, \citealt{hof14}). NOMIC has a pixel scale of 17.9\,mas/pix and the
diffraction limited single aperture point spread function (PSF) has an FWHM of 313\,mas in the $N'$ filter.

\subsection{Observing strategy}
\label{sect_obs_strategy}
To obtain a calibrated science observation, an observation of a science target (SCI) is paired with a calibrator
observation (CAL). The goal is to obtain for each science target a minimum of three such observations. Two calibrated
science observations are typically arranged in a sequence CAL1--SCI--SCI--CAL2, and two such sequences are typically
observed per science target. Ideally, four different calibrators are used. The two sequences can be observed
independently on different nights and -- if needed -- can be broken up into the original CAL--SCI or SCI--CAL pairs.
Observations of one calibrated science pointing take typically 50\,min to 1\,h.

Observations of SCI and CAL stars are carried out using the same strategy and contain $N_{\rm nods}$ nodding cycles for
background subtraction, a photometric observation, and a sky background observation. Dark frames at the target
elevation are taken during each telescope preset to a new star. During the nodding cycles the beams from both apertures
are brought to destructive interference (nulled). The optimum OPD (setpoint) is determined after each nod by minimizing
the residual $N$~band flux on target. This randomizes residual errors in the setpoint search and corrects for temporal
drifts due to atmospheric water vapor, telescope, and instrumental effects. A nod cycle consists of two on source nod
positions. In each position, $N_{\rm frames}$ frames with an integration time of typically 45\,ms per frame are
obtained. We initially set $N_{\rm nods} = 4$ and $N_{\rm frames} = 1000$, resulting in $2\times4\times1000 = 8000$
frames per observation of a star. Based on the experience with the initial reduction of our survey data, we changed
these parameters in January 2017 to $N_{\rm nods} = 3$ and $N_{\rm frames} = 2000$ ($2\times3\times2000 = 12000$ frames
per observation). At the same time, a small, stepwise phase modulation (0.2\,rad in $N$~band at a frequency of
$\sim$0.4\,Hz) has been added to break the degeneracy between null depth and residual OPD offset (imperfect destructive
interference)\footnote{An imperfect setpoint degrades the instrumental null (more stellar flux is transmitted). When
observing at a fixed OPD, this effect cannot be distinguished from actual, circumstellar emission. Modulating the OPD
during the observations eliminates this degeneracy, because the OPD dependent flux is different for the two cases.}.
The changes made in the observing strategy reduce the statistical uncertainties by a factor of $\sim$2 and increase the
observing efficiency without introducing any known systematic effects (see Sect.~\ref{sect_errors} for a discussion of
systematics). Thus, all data can be treated and analyzed in a uniform way.

If a significant fraction of the data obtained appears corrupted (e.g., phase loop opened) in our real time quality
control, additional frames or nods are taken. For the photometric observations, the two beams are separated and the
total flux of the source is measured on the two apertures independently, obtaining 500 frames (45\,ms each). Finally,
the telescope is offset to obtain 1000 sky background frames to be used for sky subtraction of the photometric frames.

\section{Data reduction and zodi measurements}
\label{sect_reduction}

\subsection{Data reduction summary}
\label{sect_reduction_summary}
Data reduction follows the strategy outlined by \citet{def16} with minor updates. After a basic reduction of each frame
(nod subtraction, bad pixel correction), the source position on the detector is determined for each nod position and
photometry (relative to the total stellar flux, measured on the photometric frames) is performed on each single frame.
The raw null depth and its uncertainty are determined using the null self calibration (NSC), a statistical calibration
method originally developed for the Palomar Fiber Nuller experiment \citep{men11, han11} and updated for the LBTI
\citep{def16, men16a}. It combines all frames recorded within a given nod.

The measurements from all calibrators in a calibration sequence are combined to determine the instrumental null depth
(nulling interferometric transfer function, TF) after correcting for the calibrator diameters. We assume a constant TF
as it is found to be stable within our uncertainties over a calibration sequence: The measurements are first filtered
to reject points for which the NSC produced a poor fit to the data ($\chi^2>5$, less than $\sim$2\% of the data), and
obvious outliers in terms of null value or uncertainty (a sign of bad background subtraction or bad data quality,
$\lesssim5\%$ of the data for a typical night). The remaining measurements from all nod positions are combined using
the error weighted mean.

The uncertainty of the final measurement has two main contributions: (1) the uncertainties of the single null
measurements obtained from the NSC fit to all data obtained in one nod, and (2) a systematic uncertainty for each nod
from imperfect background subtraction. The first contribution can be estimated by combining the NSC uncertainties of
the data obtained in each nod to the standard error of the mean. The latter uncertainty is estimated from the scatter
(root mean square) of the measurements from all nods, ignoring their NSC uncertainties. The two components are added in
quadrature.

The TF is used to calibrate the null measurements of the science target which are combined using the same strategy and
error estimation as for the calibrator observations. The uncertainty of the TF is added in quadrature to the
measurement uncertainty as an additional error term. Observations of the same science target from different nights are
combined using the error weighted mean and its standard error is derived from the uncertainties of the individual
measurements. The result is one measurement of the source null (or astrophysical null $N_{\rm as}$) and its uncertainty
per target.

\subsection{Measurement uncertainties}
\label{sect_errors}

There are no known, significant systematic uncertainties in our observations that are not already taken into account in
the above estimates. Significant general systematics in our observations can be ruled out statistically by analyzing
the null distribution of non-detections (Sect.~\ref{sect_significance}). The error from uncertain stellar diameters of
our science targets and calibrators is negligible at the LBTI's angular resolution ($<$0.01\% null depth error of
single calibrators, further reduced due to the use of multiple calibrators per science target). The risk of bad
calibrators (with companion or circumstellar emission) is minimized by using different calibrators for each science
target. No bad calibrators at our sensitivity have been identified in our observations so far. The effects of different
pointing directions between science targets and calibrators are mitigated by choosing nearby calibrators (typically
within 10 degrees), in particular with similar elevation, and are randomized by using several calibrators. We see no
significant effect of pointing direction for our selected calibrators. We also see no effects of target brightness in
any band. While the magnitude difference between calibrators and corresponding science targets can be as large as a few
magnitudes in the visible, all stars observed are by far bright enough for the AO to run at peak performance. In
$K$~band the calibrators are typically within one magnitude of the corresponding science targets, so that the effects
on OPD and tip-tilt tracking (still running at peak performance for all stars) are minimal. In $N$~band the statistical
effects of photon and detector read-out noise are dominating in addition to imperfect subtraction of the dominant mIR
background and imperfect setpoint (instrumental null), which are randomized between nods and estimated statistically as
described above. For typical observations, the dominant sources of uncertainty are the background subtraction and for
faint stars the background photon noise and detector read noise. Both can be considered statistical uncertainties in
our observations and data reduction.

\subsection{Dust distribution and zodi definition}

\begin{figure*}[t]
 \includegraphics[height=150pt, clip=true, bb = 50 -20 530 497]{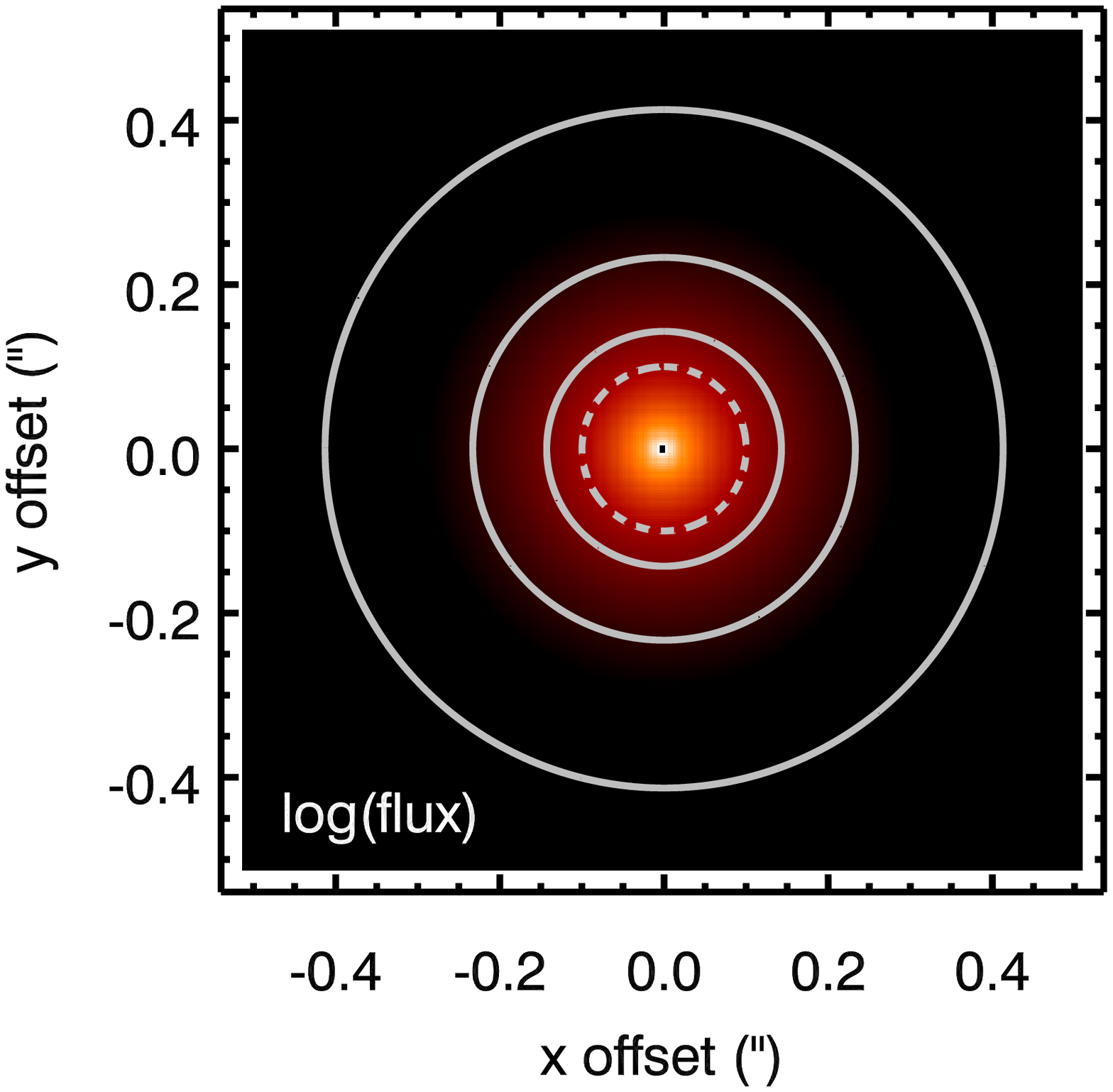}
 \includegraphics[height=150pt, clip=true, bb = 110 -20 530 497]{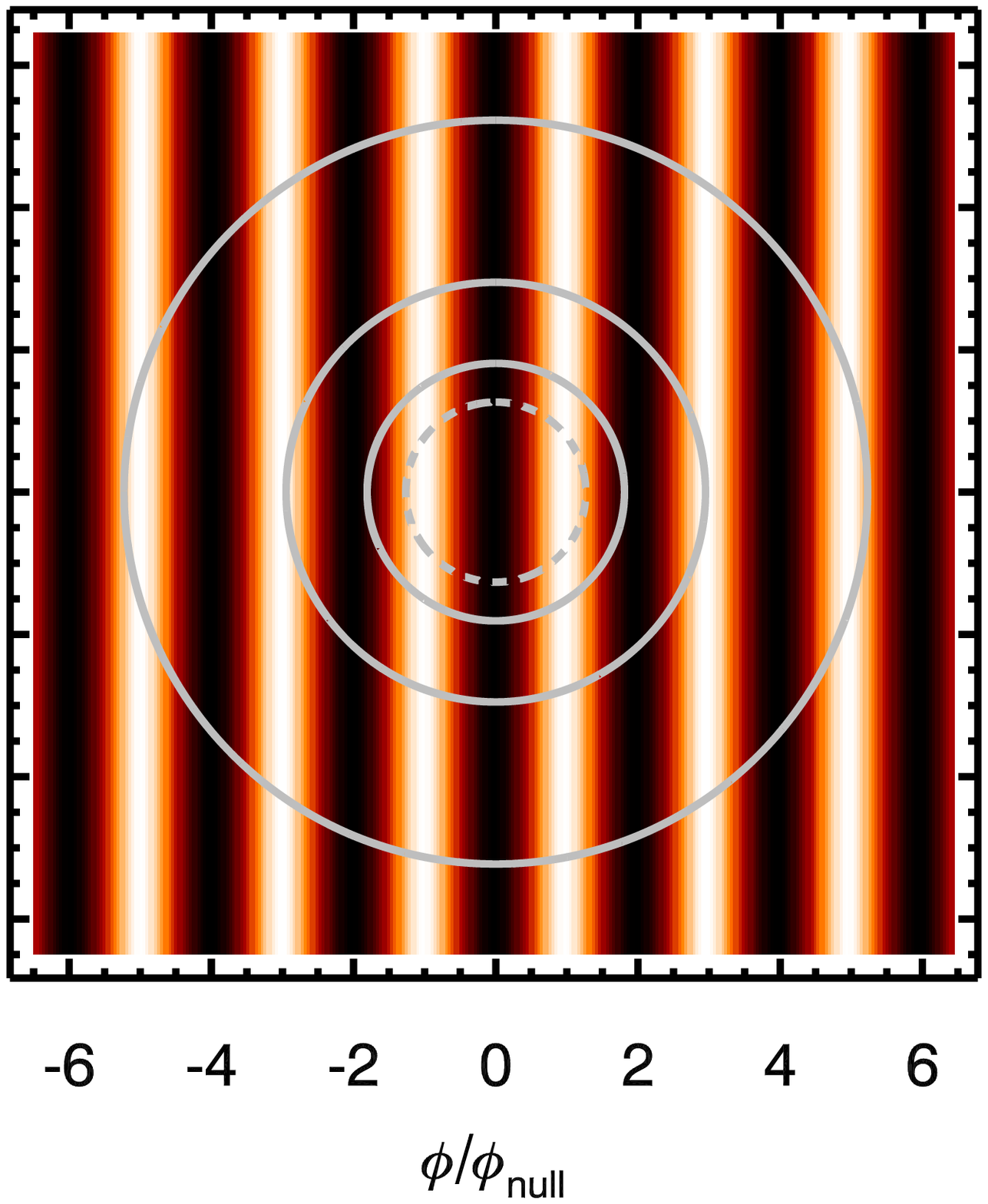}
 \includegraphics[height=150pt, clip=true, bb = 110 -20 530 497]{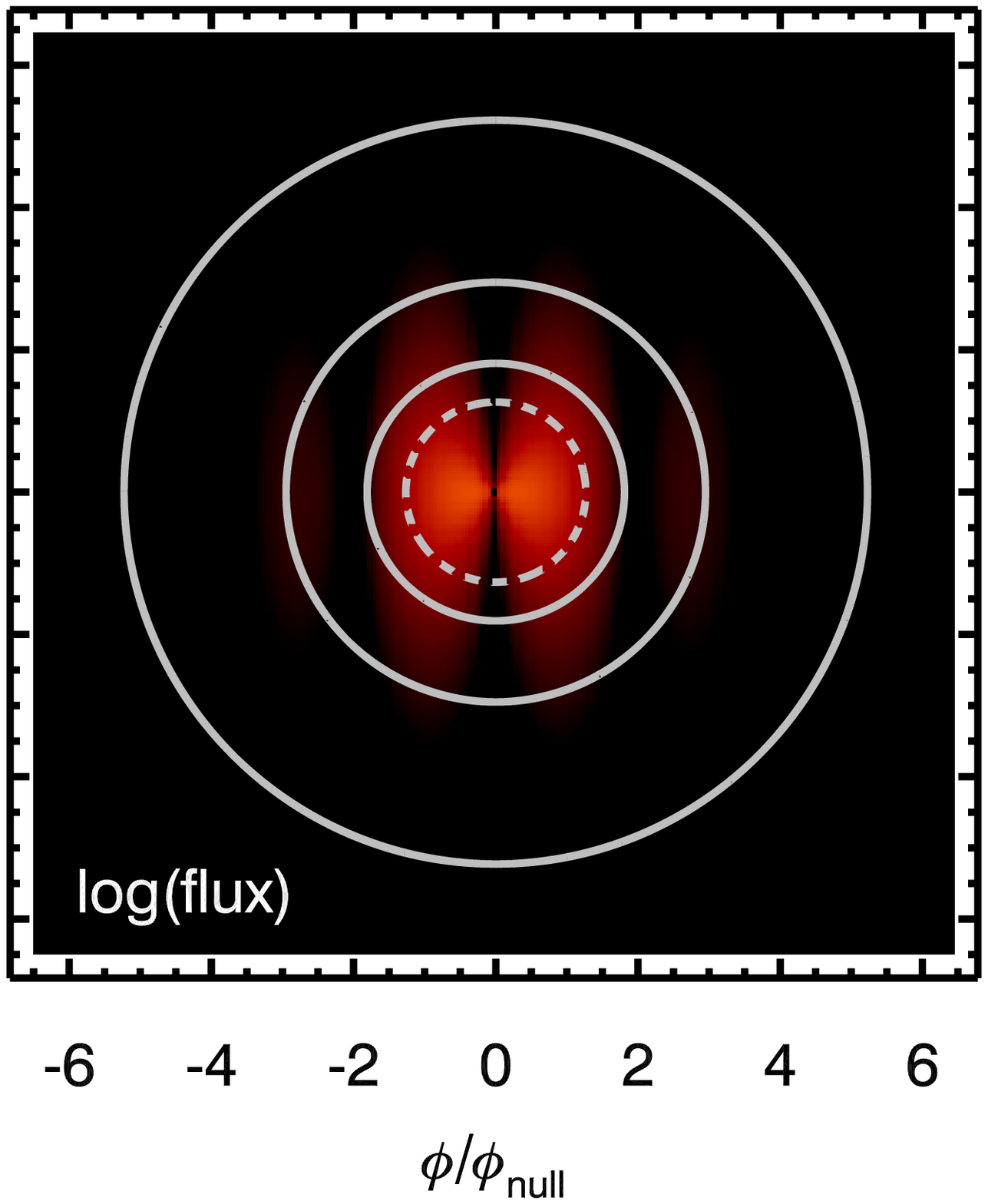}
 \includegraphics[height=150pt, clip=true, bb = 110 -20 530 497]{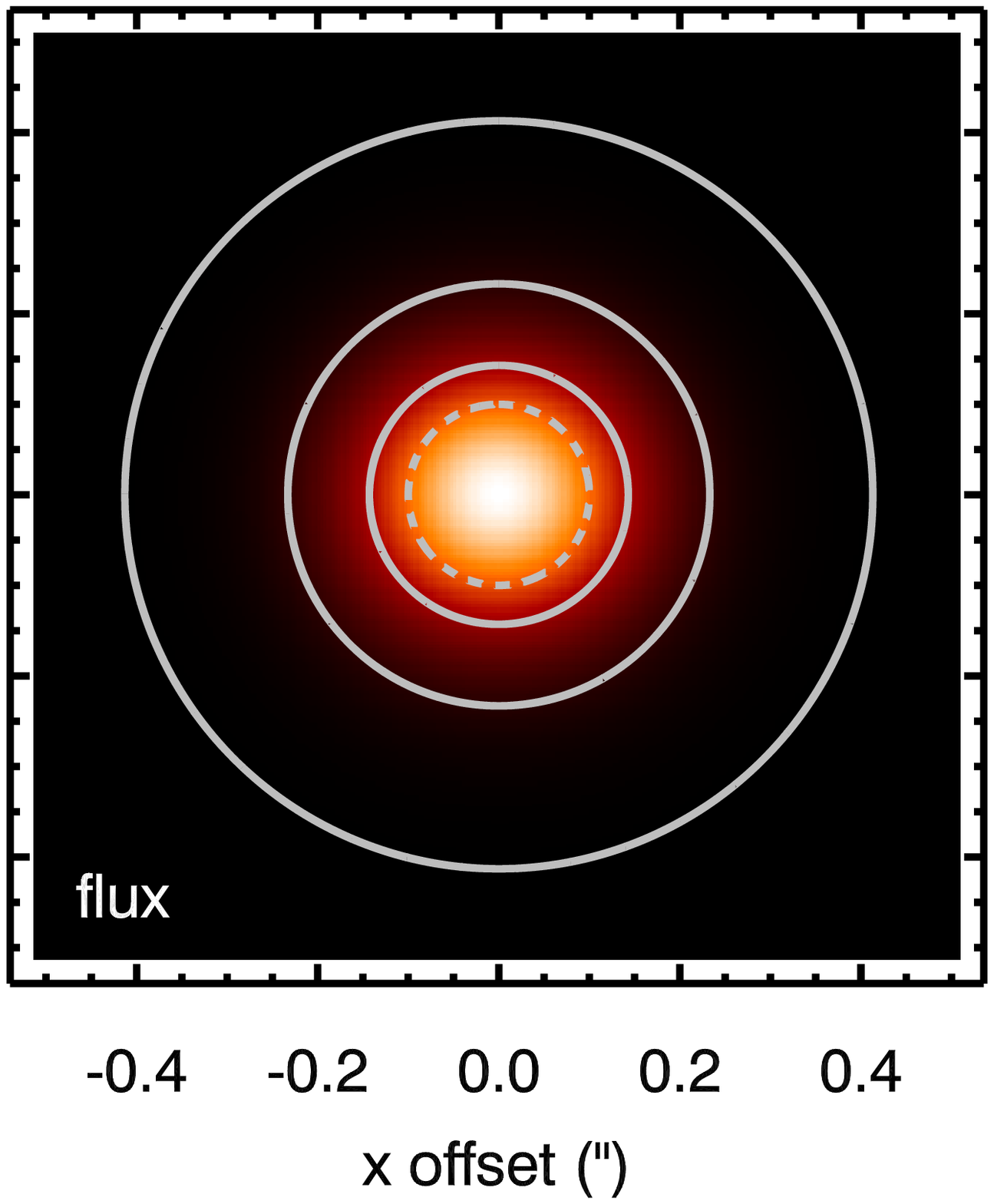}\\
 \includegraphics[height=150pt, clip=true, bb = 50 -20 530 497]{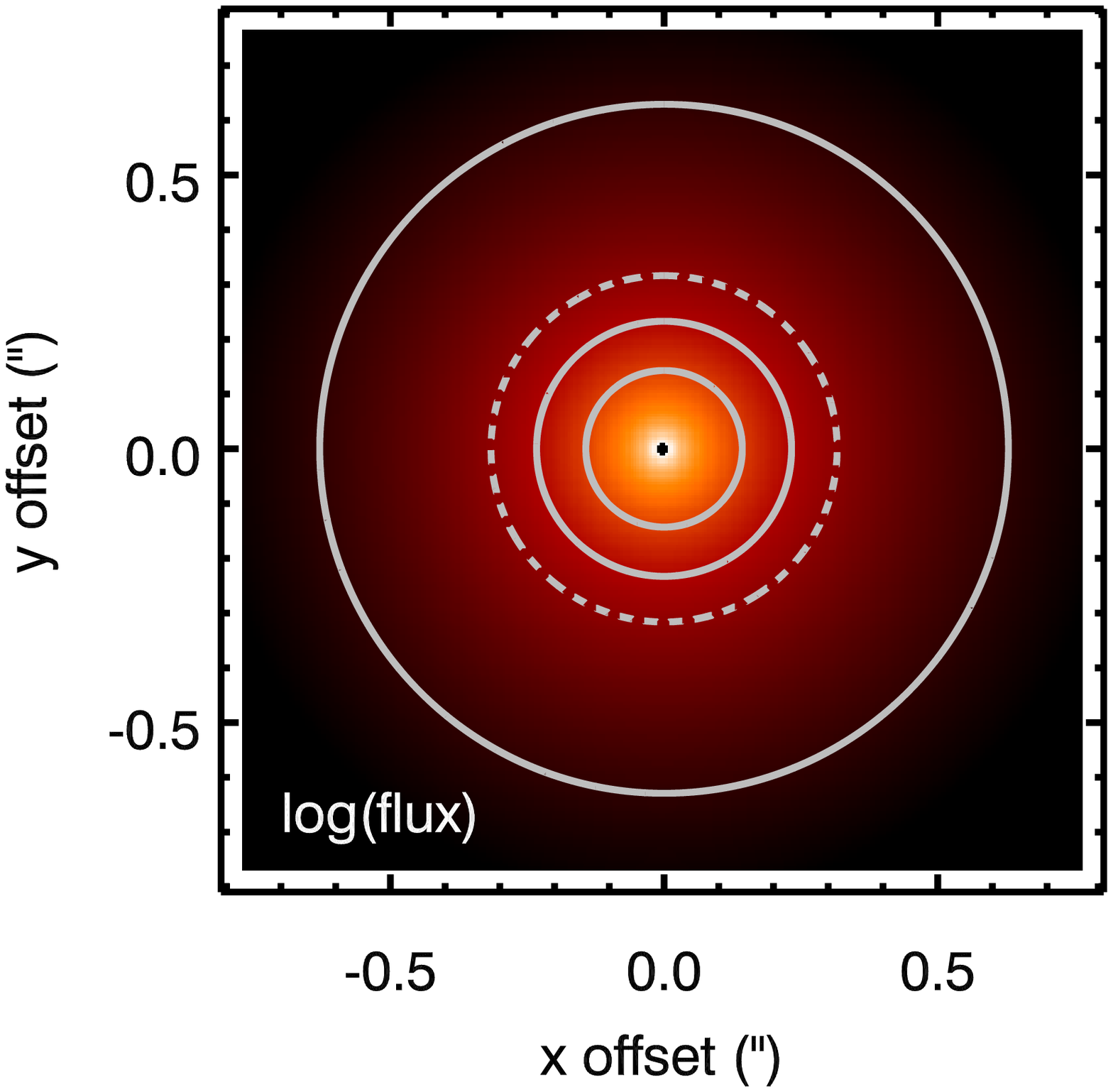}
 \includegraphics[height=150pt, clip=true, bb = 110 -20 530 497]{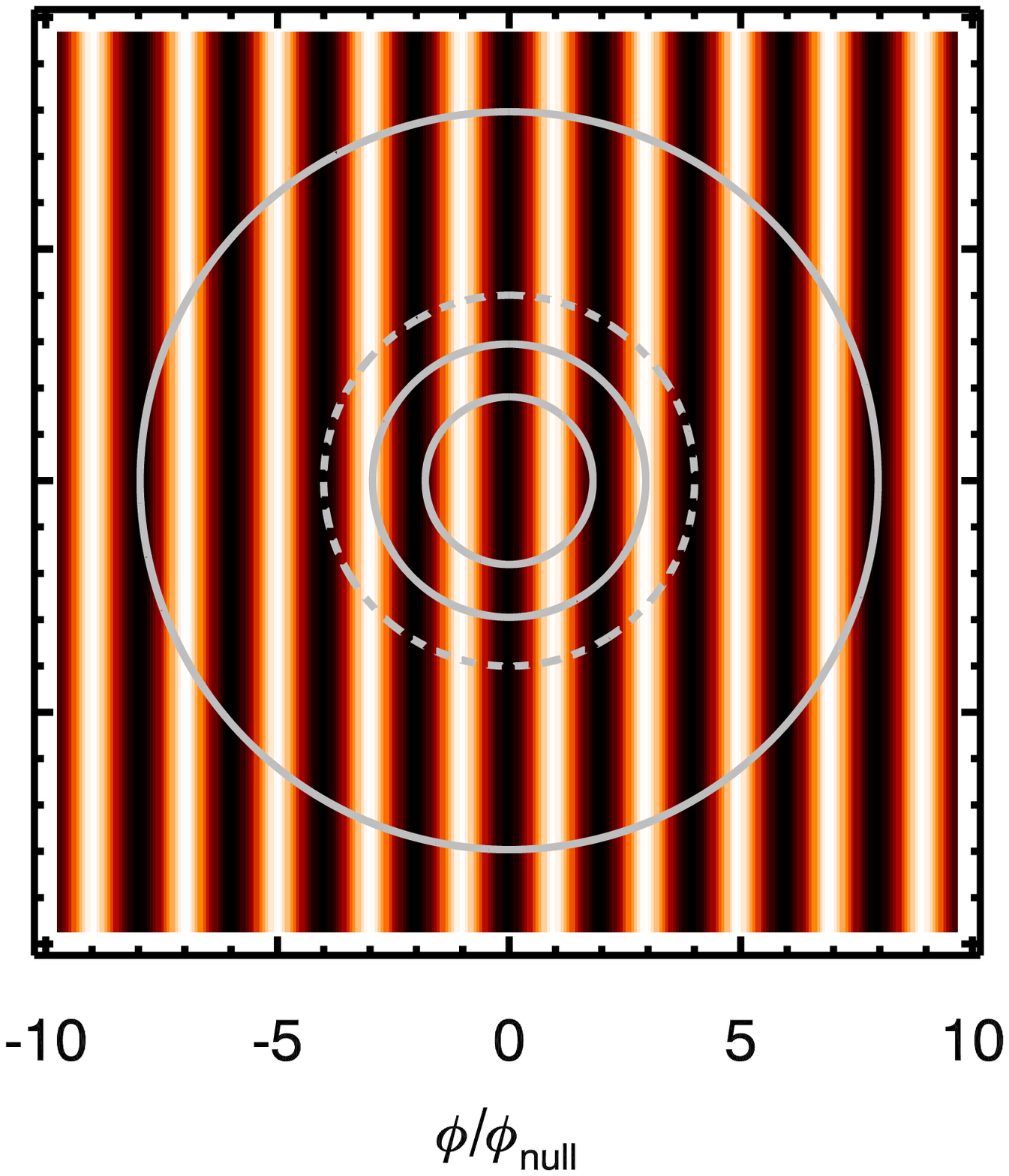}
 \includegraphics[height=150pt, clip=true, bb = 110 -20 530 497]{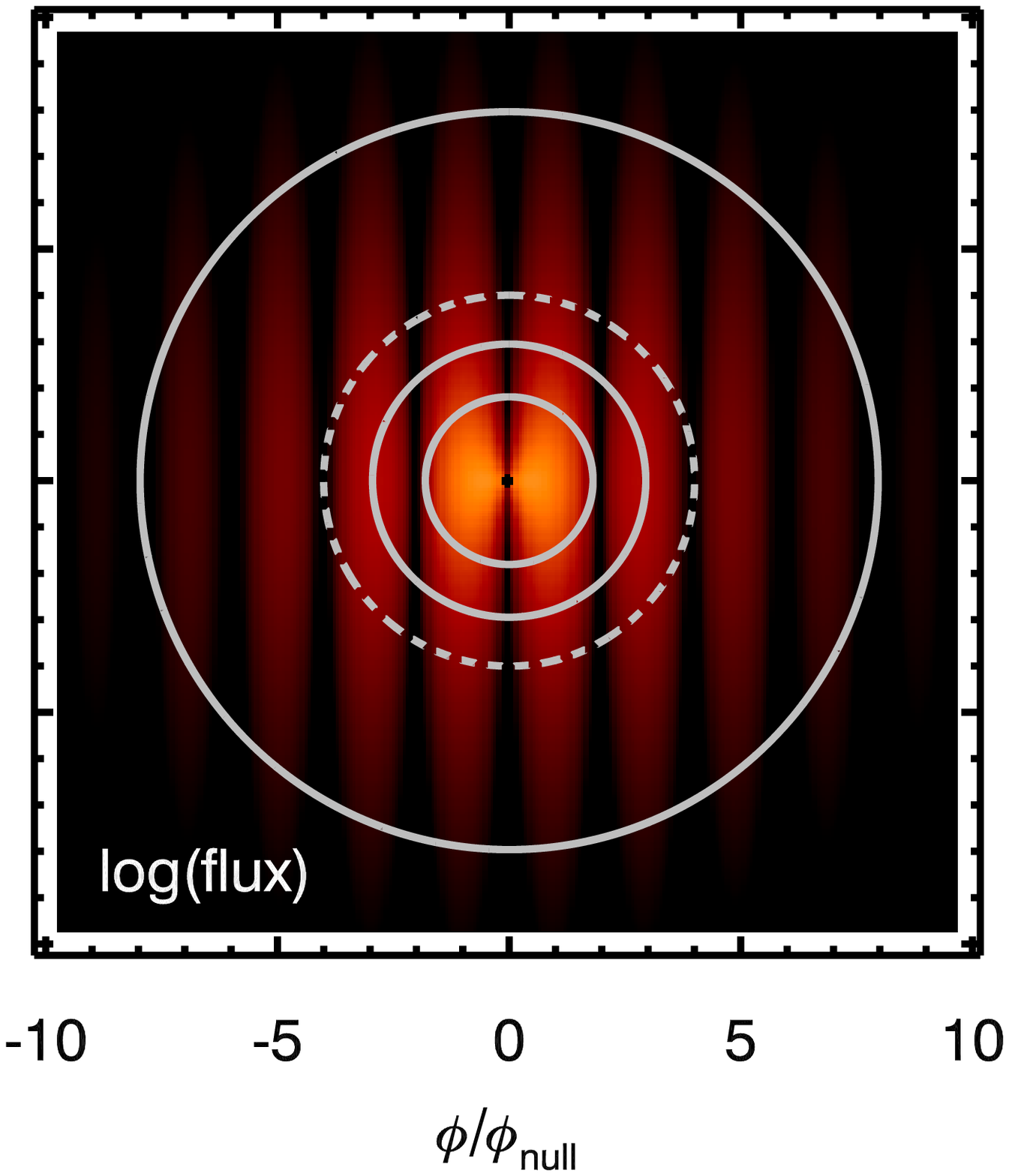}
 \includegraphics[height=150pt, clip=true, bb = 110 -20 530 497]{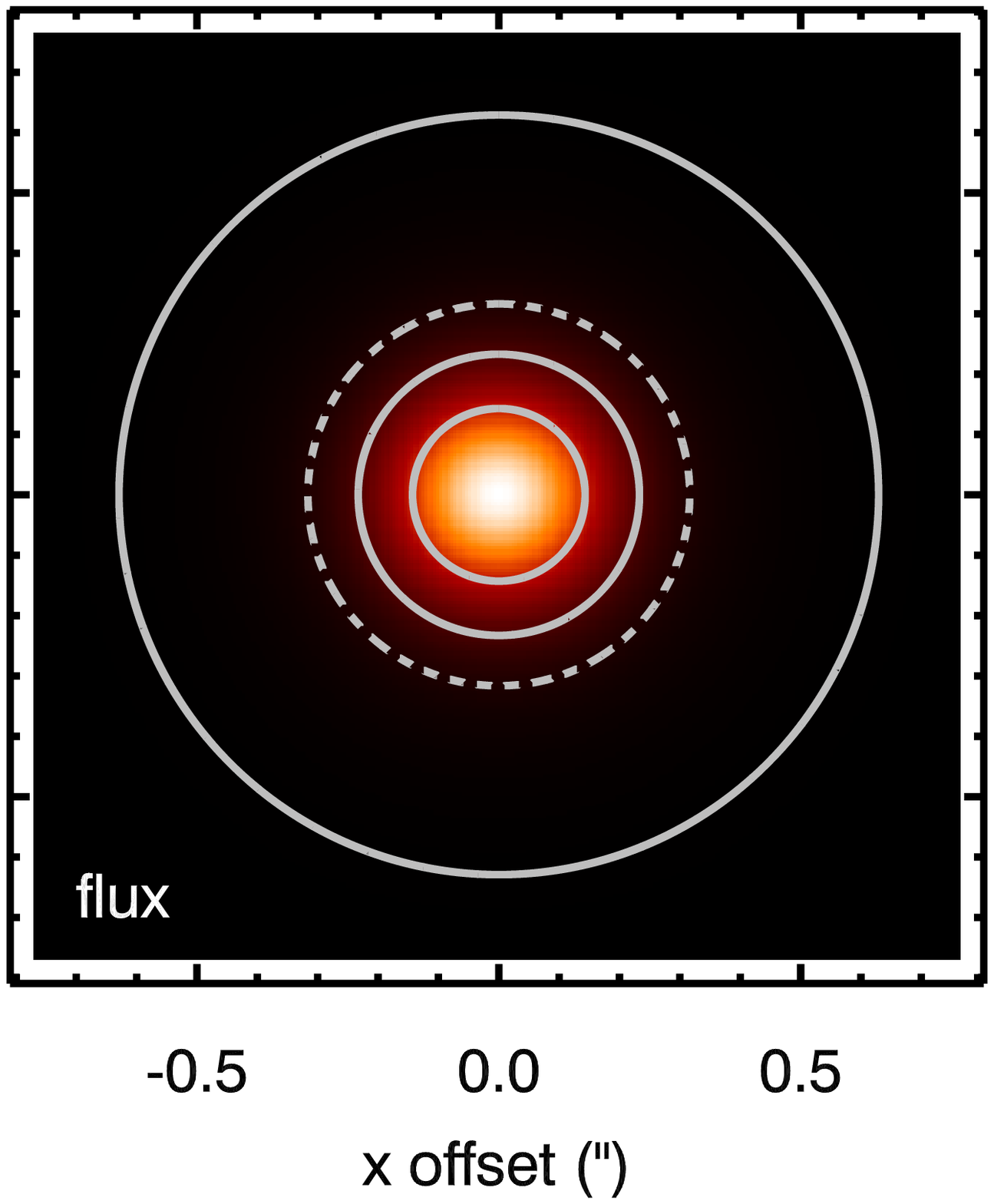}
 \caption{Illustration of the physical and instrumental scales relevant for HOSTS observations and the aperture sizes
  used in this work. Model observations are shown for two example stars (\emph{top row}: 1\,L$_\odot$, \emph{bottom
  row}: 10\,L$_\odot$) at a distance of 10\,pc. The panels show \emph{from left to right}: Our disk model for a face-on
  disk, the LBTI transmission pattern, the transmission pattern applied to the disk model, and the final simulated
  observation after convolving with the single aperture beam. Images are shown in the detector frame, so that the sky
  rotates around the star under the transmission pattern during the observations. The disk model with and without the
  transmission pattern applied are shown in logarithmic scale for better visibility, while the other images are shown
  in linear scale. The dashed circle marks the location of the EEID. The three solid circles mark from inside out the
  8\,pix, 13\,pix, and conservative apertures.}
 \label{fig_simul}
\end{figure*}

The goal of the HOSTS survey is to constrain the surface density of dust in the HZs of the observed stars. Converting a
measured null depth to surface density requires knowledge of -- or an assumption on -- the radial dust distribution,
because we are sensitive to excess originating not only in the habitable zone (measured in this work as the Earth
Equivalent Insolation Distance, EEID, the distance at which a body receives the same energy density from the star as
Earth in the Solar system), but from a range of separations from the star. The radial dust distribution can in
principle be constrained from our measurements for detected excesses with high S/N if the HZ is at least marginally
resolved by the single aperture beam of the instrument and/or if sufficient auxiliary data are available (e.g.,
\citealt{leb13, def15}). In most cases, however, this is not possible. In particular for non-detections where an upper
limit on the zodi level must be derived, we need to make an assumption for the radial dust distribution.

We assume a radial dust distribution analogous to that in our Solar system with a shallow, inward increasing dust
surface density (\textbf{Solar zodi model, SZ~model}, \citealt{kel98, ken15}). This model and its application to the
HOSTS data have been motivated and discussed in detail by \citet{ken15} and we only provide a summary here. It is
defined by a power-law radial surface density distribution with exponent $\alpha=-0.34$ and inner and outer cut-off
radii $r_{\rm in}$ and $r_{\rm out}$. We set $r_{\rm in}$ to the approximate dust sublimation distance at a black body
temperature of 1500\,K and $r_{\rm out}$ to $10\,\text{AU} \times \sqrt{L/L_\odot}$ (scaling the distance with the
square root of the stellar luminosity, so that a body at this location receives the same energy density as at 10\,AU
from the Sun). The inner edge is then small compared to the inner working angle of the LBTI ($0.5\times \lambda / B
\sim70$\,mas at a baseline $B$ of 14.4\,m) and the outer edge is large compared to the EEID. Dust inside $r_{\rm in}$
is not visible to the LBTI and dust outside $r_{\rm out}$ is cold, so that its emission in $N$~band is negligible. A
power-law is the simplest and most general description of the radial dust distribution possible. The assumption of 
$\alpha=-0.34$, in analogy to the Solar system, is a reasonable choice. Simulated images from our model in $N$~band for
face on disks around stars with luminosities of 1\,L$_\odot$ and 10\,L$_\odot$ at a distance of 10\,pc are shown in the
left column of Fig.~\ref{fig_simul}. We scale this dust distribution so that it has the same vertical geometrical
optical depth of $7.12 \times 10^{-8}$ at the EEID as the solar system zodiacal dust at a distance of 1\,AU, which
defines our unit of 1\,zodi. We compute the expected signal of this model in our LBTI observations by applying the LBTI
transmission pattern \citep{ken15} and convolving with the single aperture PSF (the resolution element in our -- nulled
or classical -- NOMIC images). This is also illustrated in Fig.~\ref{fig_simul}. Fitting the expected signal from this
model to the null measurement from a HOSTS observation of a target provides us with a measurement of the habitable zone
surface density and its uncertainty in units of 1\,zodi for each observed star. The free parameters used are the
scaling factor and the disk orientation (position angle and inclination) with respect to the hour angle range traced by
the LBTI baseline.

It is important to note that the SZ~model is a simple, geometric model motivated by the dust distribution in our Solar
system, which is the only available template. It is not necessarily a good representation of a specific exozodiacal
dust system. The relative contributions of local dust creation through asteroid collision and comet evaporation, the
effects of Poynting-Robertson (PR) and stellar wind drag, the interaction with planets, and the dust removal due to
collisions and radiation pressure blow-out are unknown for most systems. For example, the surface density of an
exozodiacal dust disk has a strong impact on its morphology as collisions will deplete dust more quickly in more dense
disks (more massive or dust accumulated in resonances with planets), while transport by PR and stellar wind drag will
dominate for more tenuous disks \citep{wya05, sta09b}. This generally leads to a stronger depletion of dust in the
inner regions of more massive disks where the orbital time scale (which are linked to the collision time scale together
with the disk's surface density, \citealt{bac93}) is shorter.

These caveats are, however, in part mitigated by the design and sensitivity of the LBTI. As can be seen in Fig.
\ref{fig_simul}, the location of the first peak of LBTI's transmission pattern is similar to the angular EEID for most
of our targets (in particular the Sun-like stars in our sample). Thus, the sensitivity of the LBTI to dust much closer
than the EEID is relatively low. Dust much further away from the star than the EEID is colder and thus fainter in the
$N$~band. As a consequence, the region in which we are most sensitive to the dust, the region of interest (the HZ), and
the location at which we normalize the dust surface density of our model are very close to each other, minimizing the
impact of the uncertain radial dust distribution. In addition, considering collisions and transport of dust grains
only, the Solar system dust distribution is best suited as a template for less dusty systems that typically result in
non-detections in our observations. For our detected, typically more massive disks, more appropriate models can be
derived from our data, follow-up observations, and a detailed analysis of auxiliary information about each specific
system in future work. In contrast, it is important for the present work to treat all systems in a uniform way.

For detected excesses, we test whether our SZ~model is consistent with the null measurements at different aperture
sizes (see Sect.~\ref{sect_opt_ap} for a discussion of the aperture sizes used). While a large range of surface density
slopes is possible for most stars due to the typically large uncertainties on the null measurements, all detections are
consistent with the SZ~model ($\alpha=-0.34$).

\subsection{Optimum aperture size}
\label{sect_opt_ap}

The residual source flux in each frame after nulling the central star is measured using aperture photometry and
calibrated using an observation of the target star with the two apertures independently (separated on the detector, no
interference). The aperture size used is a critical parameter for the result: A small aperture may not cover all the
extended dust emission, including the HZ of a system and thus omit the emission we are most interested in. On the other
hand, a larger aperture will produce a larger uncertainty due to photon noise of the sky background, detector readout
noise, and differential wavefront aberration effects between the 2 telescopes that are not captured by NSC. To find the
ideal aperture size given the expected flux distribution on our detector, we first simulate images from the SZ~model
for face-on disks (worst-case in terms of flux loss due to limited aperture size), representative of the range of
angular EEIDs around our sample stars. Two representative examples are shown in Fig.~\ref{fig_simul}.

We find that for this model the majority of the emission is always concentrated in the innermost regions. Even for the
largest EEIDs, the main emission is only marginally resolved by the single aperture NOMIC PSF. This can be explained by
the shallow, inward increasing surface density profile and the fact that dust closer to the star is warmer and thus
more luminous. However, our relatively large inner working angle of 70\,mas blocks the regions far inside the HZs of
most systems (inside 1/3 of the EEID on average for our sample). Thus, this emission is of little concern for us.
Assuming that our measurement uncertainties are dominated by background and read noise (as confirmed by null
measurements on calibrators), we find that an aperture radius of 13~pixels (233\,mas) is very close to the aperture
that yields the highest expected signal-to-noise ratio (S/N) for the SZ~model around all our targets. We thus use this
\textbf{13\,pix~aperture} as default for our null measurements. This aperture is large enough to encompasses the
physical size of the EEID directly for all Sun-like stars in our sample and only misses the part of the HZ emission in
the wings of the single aperture NOMIC PSF.

Because the actual radial dust distribution around our targets is not known, we use two more aperture sizes of
potential interest: First, an aperture radius of one half width at half maximum of the NOMIC PSF -- quantized to an
\textbf{8\,pix~aperture} of 143\,mas -- covers one angular resolution element, which is relevant in case the emission
is very compact. This could for example be the case for very hot dust, for which the LBTI could see the Rayleigh-Jeans
tail of the emission if it is located outside LBTI's central dark fringe.

Second, we use an aperture large enough to miss only negligible amounts of any significant $N$~band emission from the
system. We find that an aperture with a radius of 1\,EEID plus 1\,FWHM (full width at half maximum, 313\,mas) of the
single aperture NOMIC PSF achieves this goal. While the flux lost for any realistic dust distribution will be
negligible, this large aperture size will be particularly affected by noise. With this aperture we are thus the least
sensitive, but it is conservative in terms of neglected flux \textbf{(conservative aperture)}.

\begin{deluxetable*}{|cc|cc|cc|ccc|cccc|c|}[]
\tablecaption{Null measurements and zodi levels.\label{tab_nulls_zodis}}
\tablecolumns{14}
\tablewidth{0pt}
\tablehead{
\hline
\multicolumn{2}{r}{\textbf{Aperture $\rightarrow$}} & \multicolumn{2}{c}{\textbf{8\,pix}} & \multicolumn{2}{c}
{\textbf{13\,pix}} & \multicolumn{3}{c}{\textbf{Conservative}} & & & & & \\
\hline
HD & Name & $N_{\text{as}}$ & $\sigma_N$ & $N_{\text{as}}$ & $\sigma_N$ & $r_{\text{ap}}$ & $N_{\text{as}}$ &
$\sigma_N$ & aperture & $N_{\text{as,1}}$ & $z$ & $\sigma_z$ & $z/\sigma_z$ \\
number &  & (\%)            & (\%)       & (\%)            & (\%)       & (pix)           & (\%)            & (\%) 
& for zodi & (\%)              & (zodi) & (zodi)  &
}
\startdata
\multicolumn{13}{l}{\textbf{Sensitivity driven sample (Spectral types A to F5):}}\\
\hline
  33111   & $\beta$\,Eri    & -0.004 & 0.110 &  0.168 & 0.119 & 18 &  0.372 & 0.176 & 13\,pix & 4.45$\times$10$^{-3}$ &
    37.8 &  26.7 &  1.4 \\  
  81937   & 23\,UMa         &  0.003 & 0.073 &  0.013 & 0.092 & 25 &  0.008 & 0.179 & 13\,pix & 2.60$\times$10$^{-3}$ &
    4.9 &  35.3 &  0.1 \\
  \textbf{95418}   & \textbf{\boldmath{$\beta$}\,UMa}    &  \textbf{0.920} & \textbf{0.055} &  \textbf{1.019} &
    \textbf{0.060} & \textbf{33} &  \textbf{1.655} & \textbf{0.102} & \textbf{13\,pix} & \textbf{6.49\boldmath{
    $\times$10$^{-3}$}} &  \textbf{156.9} &   \textbf{9.2} & \textbf{17.1} \\
  97603   & $\delta$\,Leo   &  0.028 & 0.051 &  0.033 & 0.055 & 32 & -0.013 & 0.143 & 13\,pix & 5.49$\times$10$^{-3}$ &
    6.1 &  10.0 &  0.6 \\
  103287  & $\gamma$\,UMa   & -0.037 & 0.033 &  0.003 & 0.031 & 35 &  0.083 & 0.080 & 13\,pix & 7.02$\times$10$^{-3}$ &
    0.4 &   4.4 &  0.1 \\
  \textbf{106591}  & \textbf{\boldmath{$\delta$}\,UMa}   &  \textbf{0.366} & \textbf{0.094} &  \textbf{0.436} &
    \textbf{0.109} & 28 &  0.523 & 0.184 & \textbf{13\,pix} & \textbf{5.12\boldmath{$\times$10$^{-3}$}} & \textbf{85.2}
    & \textbf{21.2} &  \textbf{4.0} \\
  108767  & $\delta$\,Crv   & -0.333 & 0.131 & -0.243 & 0.199 & 26 &  0.933 & 0.365 & 13\,pix & 7.45$\times$10$^{-3}$ &
    -32.6 &  26.8 & -1.2 \\
  128167  & $\sigma$\,Boo   & -0.019 & 0.096 & -0.006 & 0.118 & 22 &  0.417 & 0.252 & 13\,pix & 2.10$\times$10$^{-3}$ &
    -2.7 &  56.0 & -0.1 \\
  129502  & $\mu$\,Vir      & -0.006 & 0.092 &  0.183 & 0.110 & 25 &  0.192 & 0.198 & 13\,pix & 1.95$\times$10$^{-3}$ &
    93.8 &  56.7 &  1.7 \\
  172167  & $\alpha$\,Lyr   & -0.037 & 0.050 &  0.022 & 0.061 & 37$^{a}$ &  0.240 & 0.150 & 13\,pix &
    4.62$\times$10$^{-3}$ &    4.7 &  13.1 &  0.4 \\
  187642  & $\alpha$\,Aql   & -0.032 & 0.166 &  0.217 & 0.192 & 47$^{a}$ & -0.995 & 0.356 & 13\,pix &
    3.84$\times$10$^{-3}$ &   56.5 &  50.1 &  1.1 \\
  203280  & $\alpha$\,Cep   & -0.301 & 0.376 & -0.233 & 0.182 & 18 & -0.075 & 0.266 & 13\,pix & 3.36$\times$10$^{-3}$ &
    -69.4 &  54.3 & -1.3 \\
\hline
\multicolumn{13}{l}{\textbf{Sun like stars sample (Spectral types F6 to K8):}}\\
\hline
  10476   & 107\,Psc        & -0.028 & 0.083 & -0.027 & 0.122 & 21 &  0.154 & 0.181 & 13\,pix & 6.36$\times$10$^{-4}$ &
    -42 &   192 & -0.2 \\
  16160   & GJ\,105\,A      &  0.228 & 0.232 & -0.227 & 0.239 & 18 &  0.538 & 0.363 & 13\,pix & 4.49$\times$10$^{-4}$ &
    -506 &   533 & -1.0 \\
  30652   & 1\,Ori          &  0.098 & 0.183 &  0.347 & 0.217 & 28 &  0.209 & 0.351 & 13\,pix & 2.27$\times$10$^{-3}$ &
    152.5 &  95.3 &  1.6 \\
  34411   & $\lambda$\,Aur  & -0.210 & 0.095 & -0.108 & 0.079 & 22 &  0.041 & 0.136 & 13\,pix & 1.16$\times$10$^{-3}$ &
    -93.3 &  68.3 & -1.4 \\
  48737   & $\xi$\,Gem      &  0.048 & 0.099 &  0.124 & 0.098 & 27 &  0.057 & 0.229 & 13\,pix & 2.20$\times$10$^{-3}$ &
    56.4 &  44.6 &  1.3 \\
  88230   & GJ\,380         & -0.111 & 0.059 & -0.077 & 0.056 & 20 & -0.189 & 0.087 & 13\,pix & 2.59$\times$10$^{-4}$ &
    -299 &   217 & -1.4 \\
  89449   & 40\,Leo         &  0.238 & 0.263 & -0.018 & 0.290 & 21 &  1.278 & 0.578 & 13\,pix & 1.51$\times$10$^{-3}$ &
    -12 &   192 & -0.1 \\
  120136  & $\tau$\,Boo     & -0.046 & 0.191 & -0.313 & 0.148 & 22 &  0.343 & 0.456 & 13\,pix & 1.50$\times$10$^{-3}$ &
    -208.3 &  98.7 & -2.1 \\
  \textbf{126660}  & \textbf{\boldmath{$\theta$}\,Boo}   &  \textbf{0.276} & \textbf{0.082} &  \textbf{0.362} &
    \textbf{0.085} & \textbf{24} &  \textbf{0.362} & \textbf{0.103} & \textbf{13\,pix} & \textbf{1.55\boldmath{
    $\times$10$^{-3}$}} &  \textbf{234.0} &  \textbf{54.8} &  \textbf{4.3} \\
  141004  & $\lambda$\,Ser  &  0.015 & 0.036 &  0.025 & 0.047 & 23 & -0.107 & 0.117 & 13\,pix & 1.20$\times$10$^{-3}$ &
    21.0 &  39.2 &  0.5 \\
  142373  & $\chi$\,Her     & -0.063 & 0.052 &  0.112 & 0.061 & 22 &  0.071 & 0.083 & 13\,pix & 1.13$\times$10$^{-3}$ &
    99.7 &  53.7 &  1.9 \\
  142860  & $\gamma$\,Ser   &  0.037 & 0.044 & -0.009 & 0.058 & 25 &  0.023 & 0.079 & 13\,pix & 1.78$\times$10$^{-3}$ &
    -4.9 &  32.4 & -0.2 \\
  \textbf{173667}  & \textbf{110\,Her}        &  0.126 & 0.096 &  0.101 & 0.115 & \textbf{24} &  \textbf{0.561} &
    \textbf{0.157} & \textbf{cons.}    & \textbf{1.98\boldmath{$\times$10$^{-3}$}} &  \textbf{283.3} &  \textbf{79.0} &
    \textbf{3.6} \\
  185144  & $\sigma$\,Dra   &  0.027 & 0.052 & -0.075 & 0.071 & 22 & -0.096 & 0.096 & 13\,pix & 8.82$\times$10$^{-4}$ &
    -85.4 &  80.7 & -1.1 \\
  215648  & $\xi$\,Peg\,A   &  0.154 & 0.121 &  0.226 & 0.167 & 23 &  0.198 & 0.214 & 13\,pix & 1.61$\times$10$^{-3}$ &
    140 &   103 &  1.4 \\
\hline
\multicolumn{13}{l}{\textbf{Commissioning targets:}}\\
\hline
  \textbf{22049}   & \textbf{\boldmath{$\epsilon$}\,Eri} &  0.037 & 0.147 &  0.206 & 0.142 & \textbf{27} &
    \textbf{0.901} & \textbf{0.269} & \textbf{cons.}    & \textbf{1.24\boldmath{$\times$10$^{-3}$}} &    \textbf{724} &
    \textbf{216} & \textbf{3.4}  \\
  \textbf{102647}  & \textbf{\boldmath{$\beta$}\,Leo}    &  \textbf{0.470} & \textbf{0.050} &  \textbf{0.420}$^{b}$ &
    \textbf{0.054} & \textbf{32} &  \textbf{1.160} & \textbf{0.333} & \textbf{8\,pix}  & \textbf{4.00\boldmath{
    $\times$10$^{-3}$}} &  \textbf{117.4} &  \textbf{12.5} & \textbf{9.4}  \\
  \textbf{109085}  & \textbf{\boldmath{$\eta$\,Crv}}     &  \textbf{4.410} & \textbf{0.350} &  \textbf{4.580}$^{b}$ &
    \textbf{0.460} & \textbf{24} &  \textbf{4.710} & \textbf{0.890} & \textbf{8\,pix}  & \textbf{1.67\boldmath{
    $\times$10$^{-3}$}} &   \textbf{2649} &   \textbf{210} & \textbf{12.6}  \\
\enddata
\tablecomments{$^{a}$~The aperture used for these targets is smaller than the actual conservative aperture due to
  limitations of the usable detector area (Sect.~\ref{sect_opt_ap}). $^{b}$~An aperture of 10\,pix instead of 13\,pix
  is used for these stars (Sect.~\ref{sect_opt_ap}).\\
  Calibrated source null levels and uncertainties are listed for the three apertures, significant excesses are
  highlighted in bold face (Sect.~\ref{sect_opt_ap}). The size of the conservative aperture depends on the star
  (distance, luminosity) and is listed for each system (column ``$r_{\rm ap}$''). For each star, the aperture used for
  the null measurement that is converted to a zodi level is listed in column ``aperture for zodi''. For non-detections,
  the 13\,pix (default) aperture is used. For detections, the aperture that produces the highest S/N is used.}
\end{deluxetable*}

Null measurements for the 8\,pix, 13\,pix, and conservative apertures are provided in this work. We limit ourselves to
these three apertures for a general and efficient analysis of the whole sample. For detailed analyses of specific
objects, a larger range of apertures is used to extract as much information as possible from the data (\citealt{def15,
def16}; Hinz et al., in prep.).

In all cases, the inner edge of the background annulus used for the photometry is set to \mbox{1\,EEID + 1\,FWHM}. Its
width is chosen to cover an area of the same size on the detector as the photometric aperture. The typically large
inner edge of the background annulus compared to the 8\,pix and 13\,pix photometric apertures avoids the HZ and
interior regions, where significant $N$~band emission might be present. At the same time, a background annulus as close
to the photometric aperture as possible minimizes the error introduced due to inhomogeneity of the background across
the detector.

We make two exceptions for the commissioning targets $\eta$\,Crv and $\beta$\,Leo for which null measurements were
already available prior to this work (\citealt{def15, def16}; Hinz et al., in prep.). These data had to be reduced very
carefully and with more human intervention due to the less standardized observing strategy and data format, and limited
data quality. This resulted in better null accuracy than our standardized data reduction can provide for these
observations. In order to use the most accurate measurements and to avoid having different, but fully consistent
measurements in the literature, we use for these stars an aperture size of 10\,pix instead of 13\,pix. Among the
apertures for which the null depths were measured previously at high precision, this aperture is the closest to the
optimum size for these stars. Instead of the conservative aperture we also use the closest aperture size measured. We
note that these exceptions are of no consequence for the conclusions of this paper, since the excess detections for
these two stars are not in question and their zodi levels are not used in what follows. Additional exceptions are
necessary for Vega ($\alpha$\,Lyr) and Altair ($\alpha$\,Aql). For both stars the conservative aperture plus the
corresponding background annulus do not fit into the usable detector area (one stripe with a size of 128\,pix~=
$2.3''$). Thus, the conservative aperture and background annulus were set to the largest possible size (Table
\ref{tab_nulls_zodis}). For Altair the difference is minor, but for Vega the largest aperture radius used is only
$\sim2/3$ of the EEID and only $\sim1/2$ of the size of the corresponding conservative aperture. Similar to $\eta$\,Crv
and $\beta$\,Leo, the exception for Vega is of little consequence for the conclusions of this paper.

Correction factors for the flux lost to a finite aperture size and null-to-zodi conversion factors for all stars
observed are derived from our SZ~model for the apertures used. For non-detections, the null measurements and
uncertainties measured using the 13\,pix aperture are converted to zodi levels and uncertainties. In case of a
detection, we use the measurement derived from the aperture that produced the highest S/N. We find that the
uncertainties on the aperture corrections and null-to-zodi conversions caused by the unknown disk orientation are
negligible compared to the measurement uncertainties for all our targets.

\section{Results}
\label{sect_results}

\subsection{Excess significance and detection threshold}
\label{sect_significance}
\begin{figure*}[t]
 \includegraphics[width=\textwidth]{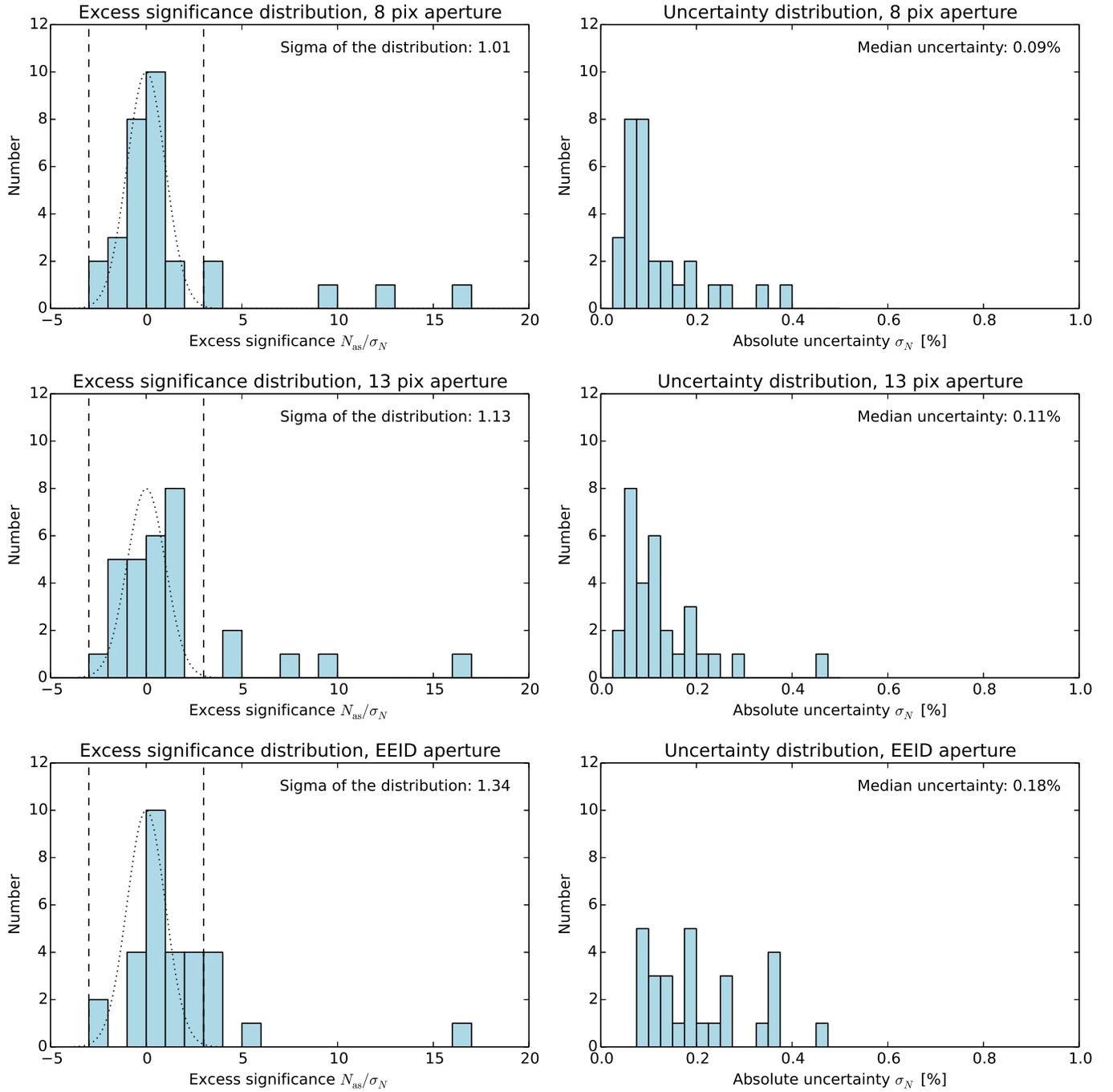}
 \caption{Distribution of excess significance $N_{\text{as}} / \sigma_N$ (\emph{left}) and uncertainties $\sigma_N$
  (\emph{right}) for our targets when measured using the three aperture sizes described in Sect.~\ref{sect_opt_ap}
  (\emph{top} to \emph{bottom}: 8\,pix, 13\,pix, and conservative). The two vertical, dashed lines in the excess
  significance distribution plots mark the $\pm 3\,\sigma$ boundaries based on our uncertainty estimates. The standard
  deviation of the distribution is computed from non-detections only ($-3 < N_{\text{as}} / \sigma_N < 3$). The dotted
  line represents a Gaussian with a standard deviation of one (Normal distribution) scaled to the peak of the histogram
  and is used only to guide the eye.}
 \label{fig_nulls_errors}
\end{figure*}

The resulting measurements and uncertainties on the source null levels ($N_{\text{as}}$, $\sigma_N$) and derived zodi
levels are listed in Table~\ref{tab_nulls_zodis} for all three aperture sizes. The distributions of the excess
significance $N_{\text{as}} / \sigma_N$ and of the uncertainties are plotted in Fig.~\ref{fig_nulls_errors}. For the
8\,pix aperture, we can see that the excess significance distribution $N_{\text{as}} / \sigma_N$ follows a Normal
distribution for $-3 < N_{\text{as}} / \sigma_N < 3$ with the addition of several measurements at $N_{\text{as}} /
\sigma_N > 3$. The distribution of the measurement uncertainty is also well behaved with the majority of the targets
having uncertainties close to the median of 0.09\% (absolute uncertainty of the null measurement, expressed as a
fraction of the total stellar flux) and a tail of measurements at larger uncertainties. This tail can in part be
explained by the fact that the observations for a fraction of our stars are still incomplete and the measurements of
the null depth of these stars are less precise.

These results validate our strategies for data reduction, null measurement, and error estimation. We thus apply a
3$\,\sigma$ threshold to identify significant excesses in our sample. We detect significant excesses around
$\beta$\,UMa, $\delta$\,UMa, and $\theta$\,Boo in addition to the two previously reported excesses around $\eta$\,Crv
\citep{def15} and $\beta$\,Leo (\citealt{def16}; Hinz et al., in prep.).

For the two larger apertures, the distribution of $N_{\text{as}} / \sigma_N$ is similar, but skewed toward positive
values. Except for the different aperture size, the approach to derive the source null levels and uncertainties is
identical for all apertures. We see no reason why an instrumental, observational, or data reduction bias should be
present when using a larger aperture but not when using the 8\,pix aperture. Effects such as an imperfect tip-tilt
correction (thus imperfect overlap of the PSFs from both apertures, resulting in less deep instrumental nulls) that
could have stronger effects at larger separation from the center of the PSFs are expected to be present for both
science targets and calibrators. Moreover, if at all significant, the effect is expected to be more pronounced on our
calibrators which are often slightly fainter in the $R$~to $I$~band than our science targets. This would skew the
distribution toward negative excesses rather than positive ones. As expected, the uncertainties increase with
increasing aperture size. While the distribution for the 13\,pix aperture is still well behaved, it is more scattered
for the conservative aperture. This is due to the variable size of the conservative apertures for each star and the
size dependent noise behavior.

Given this discussion, the shapes of the excess significance distributions for larger aperture sizes might indicate a
potential population of faint, extended excesses below our formal 3$\,\sigma$ detection threshold. These would then not
be obvious using the 8\,pix aperture, because the majority of the emission would be located outside this small
aperture. We note, however, that the changes of the histograms with aperture size are not significant and can also be
explained by statistical fluctuations due to our small sample size. As a consequence, we also employ a 3$\,\sigma$
detection threshold to identify significant excesses among the null measurements with the larger apertures. For the
13\,pix aperture we recover the three new detections made with the 8\,pix aperture at slightly larger significance, as
expected. Using the conservative aperture, we find two more detections around 110\,Her and $\epsilon$\,Eri. The fact
that no excess for these two stars was found using the smaller apertures might suggest that the dust is concentrated
close to the HZ, with a clearing closer in. However, given the large error bars, all measurements of both targets are
still consistent with the SZ~model, which also predicts a flux increase with aperture size. We claim detections for
both stars, albeit at a significance of $\sim$$3.5\,\sigma$ only in both cases, so these probable detections deserve
confirmation. The excess for $\delta$\,UMa with the conservative aperture is not significant (2.8$\,\sigma$) due to the
increased uncertainty.

\subsection{Notes on specific targets}
\label{sect_res_spec}

\noindent\textbf{\boldmath{$\epsilon$}\,Eri} (Ran, HD\,22049, HIP\,16537) is a very nearby (3.2\,pc \citealt{vanle07}),
chromospherically active K2\,V star that hosts one of the first debris disks discovered by IRAS \citep{gil86}. The disk
has been extensively studied since. The age of the star is not well known, but appears from gyrochronology and activity
calibrations to be 400~--~800\,Myr \citep{bar07, mam08}. The substantial, cool dust (fractional luminosity
$L_{\rm IR}/L_* \sim 4\times10^{-5}$, $T \sim 50$\,K, \citealt{gre14}) was first spatially resolved by sub-mm imaging
\citep{gre98} as a potentially clumpy ring at $\sim$65\,AU with a width of $<14$\,AU and a slight offset from the star
with a low eccentricity \citep{bac09, les15, macgre15, cha16, boo17}. The clumpy structure has, however, been debated
and could only be reproduced in one of several follow-up observations \citep{les15}, which suggests a contribution from
background contamination. Excess emission interior to the well-known ring has been detected by a combination of direct
imaging, mIR spectroscopy, and spectral energy distribution (SED) modeling \citep{bac09, rei11, gre14} and was resolved
with a size of $\sim$12\,AU \citep{su17}. Additional mm-wave emission at the location of the star is likely due to
stellar chromospheric emission \citep{macgre15, boo17}. We detect significant $N$~band excess emission from this star
only with the conservative aperture ($N_{\rm as} = [0.90\pm0.27]\%$). This might suggest some inner clearing of dust,
but our limited quality commissioning data are still consistent with the SZ~model for all apertures. The star was
observed with KIN, in principle at sufficient sensitivity to detect the excess we find \citep{men14}. However the use
of a pinhole limited the field of view to an FWHM of 400\,mas, more than a factor of two smaller than our conservative
aperture for this system. KIN was thus unable to detect the excess suggested by our LBTI observations. If the strong
excess is confirmed by higher quality survey data, a detailed analysis of the radial dust distribution will allow us to
put strong constraints on its origin. Our data can be compared to detailed predictions on the warm dust distribution
for various formation and evolution scenarios available in the literature \citep{mor04, bro09, rei11}. Interestingly,
the radius of the conservative aperture is close to the location of a dust clump that was tentatively detected through
$Q$~band imaging (Ertel et al., in prep.). This could indicate local dust production in the known asteroid belt and
potential shepherding by a planet interior to the belt which could also be creating the clump. There is a long history
of planet claims for $\epsilon$\,Eri, but radial velocity detection is complicated by stellar activity induced jitter.
The existence of the planet claimed by \citet{hat00} and \citet{ben06} has been debated in the literature \citep{ang12,
how16}, it is possible that a planet of period 6.8~--~7.3\,yr and mass 0.6~--~1.55\,M$_{\rm Jup}$ does orbit the star.
Attempts to infer the presence of outer planets based on the ring structure are problematic due to the uncertain nature
of the intrinsic disk morphology.

\noindent\textbf{110\,Her} (HD\,173667, HIP\,92043) is a well studied F6\,V star at 19.2\,pc. It has solar or slightly
sub-solar metallicity and an age of $\sim$3\,Gyr (compilation by \citealt{tri08}). A faint far-infrared (fIR) excess
was first suggested by \citet{eir13}. An unresolved 3~--~4$\,\sigma$ excess was confirmed by the focussed analysis of
this system by \citet{mar13} and attributed to a regular, Kuiper belt-like debris disk, although the low significance
detection may be questioned \citep{gas13}. In addition, a marginal, extended excess emission was found by both
\citet{eir13} and \citet{mar13}. It may be attributed to the presence of a very cold disk \citep{eir11, kri13}, but it
has been shown that background contamination is a likely alternative \citep{mar13, gas14}. The excess was not detected
independently by Spitzer, but its reality at 70\,$\mu$m is supported at the 2.3\,$\sigma$ level \citep{gas13}. No
excess was found at 8~--~30\,$\mu$m with Spitzer/IRS \citep{bei06, law09}. We here follow the focussed analysis by
\citet{mar13} and consider the star a debris disk host star, but urge caution with this conclusion due to the low
significance detection. \citet{abs13} and \citet{nun17} find a significant $K$~band excess using CHARA/FLUOR
interferometric observations. Similar to $\epsilon$\,Eri, we detect a mIR excess around 110\,Her only in the
conservative aperture ($N_{\rm as} = [0.561\pm0.157]\%$), but the measurements with all apertures are also consistent
within the uncertainties with the ZS~model. The absence of a massive, cold dust belt puts in question whether the HZ
dust can be produced in a (tenuous) outer disk and migrate inward due to PR~drag, as was suggested by \citet{ken15b}.
It has also been shown that the nIR excess cannot be explained by PR~drag, even in the presence of a massive, cold dust
belt \citep{wya07b, vanlie14}. Instead, both the nIR and mIR excess in this system could be produced by evaporating
comets that would not necessarily originate in a detectable Kuiper belt \citep{bon14, far17}. The star has no detected
radial velocity planet at the level of few 10\,m/s \citep{fis14, how16}.

\noindent\textbf{\boldmath{$\beta$}\,UMa} (Merak, HD\,95418, HIP\,53910) is an early A~type star at a distance of
24.5\,pc \citep{vanle07}. Excesses are consistently detected at wavelengths $>24\,\mu$m with Spitzer/MIPS at 24~and
$70\,\mu$m, and with Spitzer/IRS between 30~and $34\,\mu$m \citep{che06}. The circumstellar emission was also resolved
by Herschel/PACS at 70~and $100\,\mu$m \citep{boo13}, with a very clean fit to the data obtained by a narrow dust ring
at about 43\,AU, and seen close to edge on (inclination $\sim$$84^\circ$, position angle $114^\circ$, however with
considerable uncertainties). A warm disk component was first resolved at $11.2\,\mu$m \citep{moe10} with an inclination
and position angle consistent with that of the outer disk. Its excess flux was first quantified with KIN \citep{men14}.
We re-detect the excess at very high significance of $[0.920\pm0.055]\%$, $[1.019\pm0.060]\%$, and $[1.66\pm0.10]\%$ in
the 8\,pix, 13\,pix, and conservative apertures, respectively. A 3$\,\sigma$ upper limit of 0.43\% was put on the hot
excess around this star in $K$~band by \citet{abs13}. An origin of the HZ~dust in the massive outer disk seems
plausible for this system. The combined data from KIN and LBTI at different inner working angles, fields of view/
aperture sizes, and baseline orientations combined with the nIR and mIR to fIR spectrophotometric data and resolved
images constitute a rich data set and detailed modeling of the system by our team is ongoing.

\noindent\textbf{\boldmath{$\beta$}\,Leo} (Denebola, HD\,102647, HIP\,57632) is an A3\,V star with a luminosity of
$15\,L_\odot$, distance of 11.0\,pc, and isochronal estimates of its age ranging from $50$\,Myr \citep{che06} to
$100$\,Myr \citep{vic12}. It has been proposed to be a member of the $\sim$40\,Myr-old Argus moving group
\citep{zuc11}. $\beta$\,Leo has been identified as a probable $\delta$\,Scuti variable \citep{fro70}. As a nearby young
star it is a prime target for direct imaging campaigns, but they have been thus far unsuccessful \citep{mes15, dur16}.
The dust around $\beta$\,Leo has been studied extensively, with infrared excesses previously reported from cold dust
using IRAS \citep{aum85}, and Spitzer \citep{su06, che06}, from warm dust at $8.5\,\mu$m using KIN \citep{men14}, and
from hot dust using the CHARA/FLUOR interferometric observations in K band~\citep{ake09, abs13}. Direct imaging at
$100\,\mu$m and $160\,\mu$m with Herschel resolved a roughly face on, cold dust disk extending to $\sim$40\,AU
\citep{mat10}. The resolved disk cannot produce the observed flux at short wavelengths, leading \citet{sto10} to
produce a two component dust model with planetesimal belts from 2~--~3\,AU and 5~--~55\,AU, and \citet{chu11} to
suggest a three component model with belts at 2\,AU, 9\,AU, and 30~--~70\,AU. The LBTI detection of an excess
($N_{\rm as} = [0.470\pm0.050]\%$ in the 8\,pix aperture) and its aperture size dependence are being studied in detail
by Hinz et al.\ (in prep.), finding that the measurement is consistent with dust spiraling in from the outer belt due
to PR~drag. The origin of the hot dust remains unclear in this scenario, but could be related to comet delivery from
the outer disk.

\noindent\textbf{\boldmath{$\delta$}\,UMa} (Megrez, HD\,106591, HIP\,59774) is an A2\,V star at a distance of 24.7\,pc.
The star is a rapid rotator which has been taken into account for its age estimate of 400\,Myr \citep{jon16}. An
infrared excess at both 24~and $70\,\mu$m was identified by \citet{su06}, but only at $\sim$$4\,\sigma$ in both bands.
A re-analysis of the Spitzer data and newer Herschel data disprove the excess \citep{gas13, thu14}. Our excess
detection ($N_{\rm as} = [0.436\pm0.109]\%$ in the 13\,pix aperture) is particularly interesting, because this is a
star without any previously known dust (despite sensitive fIR observations), but with a mIR interferometric detection
of exozodiacal dust.

\noindent\textbf{\boldmath{$\eta$}\,Crv} (HD\,109085, HIP\,61174) is an early type star (spectral type F2\,V) at a
distance of 18.3\,pc. It has an age of 1~--~2\,Gyr (e.g., \citealt{ibu02, mal03, vic12}). The star is well known for
its prominent debris disk at 152\,AU and massive warm dust system \citep{wya05b, bei06, che06, lis12, duc14, mar17b}.
The latter has been studied in detail by \citet{leb16} using mid-infrared interferometry from KIN \citep{smi09, mil11}
and LBTI \citet{def15}, and available mIR and fIR spectroscopic and photometric data. We list for this star the source
null level of $N_{\rm as} = [4.41\pm0.35]\%$ published by \citet{def15} which has been measured for the 8\,pix aperture
and was found not to depend significantly on aperture size. The excess measured by the LBTI has been found to be small
compared to that expected from earlier spectroscopic and photometric observations. This, together with the aperture
independent excess led to the conclusion that the majority of the emission must be rather compact, located at a
separation of 0.5~--~1.0\,AU from the star, close to the inner working angle of the LBTI. Such large amounts of warm
dust close to a Gyr old star must be transient \citep{wya07b}, suggesting that it was produced in a recent,
catastrophic collision. Surprisingly, no nIR excess was detected around this star with a 3$\,\sigma$ upper limit of
1.5\% \citep{abs13}.

\noindent\textbf{\boldmath{$\theta$}\,Boo} (HD\,126660, HIP\,70497) is a F7\,V star at 14.5\,pc, with a luminosity of
4\,$L_\odot$, of nearly solar abundance and age of 1~--~2\,Gyr \citep{mon16}. It has been searched for infrared excess
with both Spitzer \citep{tri08} and Herschel \citep{mon16}, with no detection of excess at wavelengths up to
160\,$\mu$m. The star has not been searched for the presence of hot dust using nIR interferometry. We find an excess of
$N_{\rm as} = [0.362\pm0.085]\%$ in the 13\,pix aperture, making this star the second one -- after $\delta$\,UMa -- in
our sample for which we find a detection without previously known fIR excess.

\noindent\textbf{\boldmath{$\alpha$}\,Lyr} (Vega, HD\,172167, HIP\,91262) has, of course, the prototypical debris disk
\citep{aum84}. The star is conventionally classified as of type A0\,V. However, it is very rapidly rotating and seen
pole-on, and hence has a large temperature gradient ($\sim$2000\,K) from its pole to equator \citep{gul94, pet06,
auf06}. Vega is at a distance of $7.68\pm0.02$\,pc. The rapid rotation complicates conventional estimates for its age
\citep{pet06, yoo08}; correcting for its effects, the age is estimated at $455\pm13$\,Myr \citep{yoo10}. The Vega
debris disk is complex.  It has an excess most likely due to very hot dust in the near infrared, at a level of
$1.29\pm0.19\%$ of its photosphere at $2.12\,\mu$m, within a field of view of $\sim$7.8\,AU, and detected at a similar
level but at lower significance, i.e., $1.23\pm0.45\%$, at $1.65\,\mu$m \citep{abs13, def11}. Previous nulling
interferometry at $10\,\mu$m \citep{liu04c} established an upper limit requiring the spectrum of this excess to be as
blue as Rayleigh-Jeans, or its source to lie within 0.8\,AU of the star. KIN measurements by \citet{men14}, rule out
red excess generated between 0.05 and 1.5\,AU with a 3$\,\sigma$ upper limit of about 1.25\% of the photosphere
($\sim2.5$ times the source null). The LBTI upper limits of 0.12\%, 0.2\%, and 0.7\% source null in an aperture of
1.1\,AU, 1.8\,AU, and 5.1\,AU (8\,pix, 13\,pix, and conservative aperture), respectively, with an inner working angle
of 0.5\,AU further strengthen the constraints on the dust location and emission. Considering all arguments, the hot
dust is most likely located inside LBTI's inner working angle and the KIN measurement still provides the strongest
constraint on its mIR emission. This is plausible as the blackbody equilibrium temperature at 0.5\,AU from the star
would still be only $\sim$1000\,K, cool enough for most dust species to exist at this location or closer in. It appears
that its spectrum is steeper than Rayleigh-Jeans between 2 and 10\,$\mu$m (or its emission is variable). Such spectra
can be generated by very small grains of carbon, or of some of the robust oxides such as FeO expected to be produced in
the destruction of silicate grains \citep{rie16, kir17}. Another perspective on the HOSTS result is provided by the
measurements with Spitzer and Herschel as analyzed by \citet{su13}. They find that the debris spectral energy
distribution indicates an asteroid-analog belt centered at $\sim$ 14 AU from the star. The LBTI measurement shows that
the region interior to this belt must be largely devoid of dust, strengthening earlier mIR spectroscopic results
\citep{su13}. Given the strong dust emission in the nIR very close to the star, and in the mIR to fIR further away, the
void of dust at a few AU from the star is particularly curious and a clearing mechanism such as the presence of a
planet might be required to explain it.

\section{Discussion}
\label{sect_discussion}

In this section we present a statistical analysis of the observations presented in this work. In
Sect.~\ref{sect_res_detrate} we derive and analyze basic detection statistics. We then constrain in
Sect.~\ref{sect_res_medianzodi} the median zodi levels for relevant groups of stars. In Sect.~\ref{sect_lum_funct} we
combine our observations with previous work to illustrate how detailed, future modeling can constrain the exozodi
luminosity function.

\subsection{Detection rates among different subsamples}
\label{sect_res_detrate}

To derive statistics from our observations, we first define relevant subsamples of our target stars. We exclude
$\eta$\,Crv and $\beta$\,Leo from the statistical analysis. While being part of the unbiassed HOSTS target list, they
were selected as commissioning targets specifically for their known $N$~band excesses. They thus cannot be considered
part of a statistically unbiased target selection before the majority of the HOSTS targets have been observed.
$\beta$\,UMa, despite its previous KIN detection, went through our real time target selection during the observations
and can be considered an unbiased target. Our sample can be divided relatively evenly into early type stars (spectral
types F5 and earlier, 12 stars) and Sun-like stars (spectral types F6 and later, 16 stars). In addition, \citet{men14}
found from their KIN results that the detection rate of warm dust is higher for stars with previously known cold dust
than for stars without. While with $\delta$\,UMa and $\theta$\,Boo we find the first two cases of mIR interferometric
excesses without previously known dust, our detection rates seem generally consistent with this conclusion. They also
find a tentative anti-correlation between nIR detected hot dust and their KIN detections. Because this was based on
very small number statistics and our relevant sample is similarly small, we ignore the presence of hot dust for now and
will discuss the hot dust systems separately at the end of this section. We thus divide our sample into early type and
late type stars and into stars with previously known cold dust (`cold dust stars') and without (`clean stars'). Our
derived occurrence rates of HZ dust at the sensitivity of the observations presented in this paper, and their binomial
uncertainties, are listed in Table~\ref{tab_det_rates}. To test whether the differences in detection rates measured
from different subsamples are statistically significant, we perform Fisher's exact test (Table~\ref{tab_fisher}).

Most of our detection rates are consistent with each other given the statistical uncertainties from the limited sample
sizes. However, we can rule out with a formally high confidence (probability 3\%) that the occurrence rate of HZ dust
is the same among stars with and without cold dust\, confirming the result by \citet{men14}. The result remains the
same for Sun-like stars only, but no constraints can be put on early-type stars alone (41\% probability).

\begin{table}[]
\caption{Subsamples, excess detections, and occurrence rates\label{tab_det_rates}}
\centering
\begin{tabular}{cccc}
\hline
\hline
 ~~~~~~~~~~~~ & ~~~Cold dust~~~ & ~~~~~Clean~~~~~ & ~~~~~~All~~~~~~ \\
\hline
 Early                & 1 of 3              & 1 of 9            & 2 of 12            \\[-4pt]
 ~~type               & $33^{+28}_{-15}\%$  & $11^{+18}_{-4}\%$ & $17^{+15}_{-6}\%$  \\[4pt]
 Sun-                 & 2 of 2              & 1 of 14           & 3 of 16            \\[-4pt]
 ~~like               & $100^{+0}_{-46}\%$  & $7^{+13}_{-2}\%$  & $19^{+13}_{-6}\%$  \\[4pt]
 \multirow{2}{*}{All} & 3 of 5              & 2 of 23           & 5 of 28            \\[-4pt]
                      & $60^{+16}_{-21}\%$  & $8^{+10}_{-3}\%$  & $18^{+9}_{-5}\%$   \\
\hline
\end{tabular}
\end{table}

\begin{table}[]
\caption{Probability that two samples are drawn from the same distribution\label{tab_fisher}}
{\centering
\begin{tabular}{ccc}
\hline
\hline
 ~~~~~~Samples 1~~~~~~ & ~~~~~~Sample 2~~~~~~ & ~~Probability~~ \\
\hline
 All early type   & All Sun-like     & 0.38 \\
 All dusty        & All clean        & 0.03$^{~a}$ \\
 Clean early type & Dusty early type & 0.41 \\
 Clean Sun-like   & Dusty Sun-like   & 0.03$^{~a}$ \\
 Clean early type & Clean Sun-like   & 0.50 \\
 Dusty early type & Dusty Sun-like   & 0.30 \\
\hline
\end{tabular}}
\tablecomments{$^{a}$~These probabilities are significantly affected when considering 110\,Her a clean star
 (Sect.~\ref{sect_res_spec}). In this case, the probability changes to 0.12 for `All~dusty' vs.\ `All~clean' and to
 0.19 for `Clean~Sun-like' vs.\ `Dusty~Sun-like'.}
\end{table}

Also interesting is the comparable detection rate for Sun-like and early type stars, independent of the presence of
cold dust. For Sun-like stars, our sensitivity in terms of zodi level is on average $\sim$4~times worse than for early
type stars. Thus, a similar detection rate for Sun-like and early type stars suggests a higher average dust level for
Sun-like stars (but note the cautions in the following paragraph). If confirmed, this might imply that the transport of
material to/through the HZ from further out in the system is more efficient for Sun-like stars than for early type
stars. In particular, this could mean that Sun-like stars might harbor significant amounts of HZ dust even if no
detectable amounts of cold dust are present, a conclusion also suggested by our detection on $\theta$\,Boo (although
our $\delta$\,UMa detection suggests that similar cases may exist for early type stars as well). Such a scenario would
complicate the target selection for future exo-Earth imaging missions.

It is important to note that our results are so far based on few detections mostly in the 3~--~5$\,\sigma$ range and
affected by small number statistics. They thus require confirmation from a larger sample and more sensitive
observations. In addition, the difference in detection rates between dusty and clean stars relies also on our ability
to identify cold dust detections. For example, considering the detection of cold dust around 110\,Her as spurious due
to background contamination and an underestimation of the measurement uncertainties would move this star with an LBTI
detection to the clean stars sample. In this case, there is no significant difference in detection rates between clean
and dusty stars with a probability of 0.12 that the two samples are drawn from the same occurrence rate. Furthermore,
the detections around $\delta$\,UMa and $\theta$\,Boo (both clean stars) demonstrate that limiting the target list of
exo-Earth imaging surveys to stars without cold dust does not guarantee that all targets have low HZ dust levels,
although we still find a lower detection rate around clean stars than around dusty ones.

Given the apparent correlation of HZ dust and cold dust, we need to exclude cold dust stars when searching for a
correlation with the presence of hot dust. This limits our available sample of hot dust systems to only three stars,
none of which shows any sign of excess related to emission close to the star in our observations. However, the small
sample size prevents any conclusion on the correlation between hot dust and HZ dust. The discussion on Vega in
Sect.~\ref{sect_res_spec} presents our strongest constraints that can be put on the hot dust systems from the available
LBTI data without detailed modeling.

\subsection{Median zodi level}
\label{sect_res_medianzodi}

The main goal of the HOSTS survey is to determine the typical HZ dust levels around nearby stars. Here, we perform a
statistical analysis of the HOSTS targets observed so far in order to provide the strongest constraints possible at the
moment. We follow the approach presented by \citet{men14} to fit a probability distribution of the zodi levels (exozodi
luminosity function) for our observed stars to our measurements using a maximum likelihood estimate. We assume a
lognormal distribution for a given star to have a specific zodi level $z$:
\begin{equation}
 p(z) = \frac{1}{z \varsigma \sqrt{2\pi}} \exp\left(-\frac{\left(\ln z - \mu\right)^2}{2\varsigma^2}\right).
\end{equation}
The likelihood of finding the measured zodi levels for a sample of stars -- given the assumed luminosity function and
the individual uncertainties on each star -- is computed for an equally spaced grid of values for the lognormal
parameters $\mu$ and sigma parameter $\varsigma$. To derive the probability for the median of the fitted distribution,
$m=\exp\left(\mu\right)$, and $\varsigma$, we then extend the approach used by \citet{men14} by performing a Bayesian
analysis. We apply a $1/m$ prior, equivalent to assuming a flat prior in $\mu$, marginalize the likelihood distribution
over $\varsigma$, and compute the posterior cumulative probability distribution function (CPDF) of $m$. From this we
can directly derive constraints on the median zodi level of our best-fit distribution for a sample of stars at any
confidence level.

The alternative and more naive approach to derive a median of our measurements directly, rather than from a fit of
the underlying distribution, does not yield good statistical results. For such an analysis one would need measurements,
rather than upper limits, for all stars considered. This would force us to ignore our non-detections, which are most
constraining for the underlying distribution, and to use only our few detections, which represent only the most extreme
cases of the probability distribution. The use of a lognormal distribution is motivated by the fact that it has a well
defined median and a small number of parameters. We also tested the other distributions used by \citet{men14}, i.e., a
uniform distribution with an upper cut-off value and a truncated Gaussian distribution. Both produce median zodi levels
that are generally consistent with those form the lognormal distribution, while the uniform distribution does not
reproduce our date well. In Sect.~\ref{sect_lum_funct}, we combine our results with available photometric constraints
and compare them to a physical model of a luminosity function, that predicts a power-law distribution \citep{ken13}.

We perform this analysis on the zodi measurements obtained with the 13\,pix and conservative apertures. The
conservative apertures are used in addition to the formally more sensitive 13\,pix aperture in order to test if
potentially neglecting a fraction of the flux using the 13\,pix aperture has any effect on our results. We still use
the SZ~model to convert null levels and uncertainties to zodi levels. We do not list the results in
Table~\ref{tab_nulls_zodis}, because this is only a sanity check, and the zodi levels derived from this are not to be
considered our final results. We will show below that the results from the 13\,pix and conservative apertures are fully
consistent.

As discussed before, we find a higher detection rate for stars with cold dust than for stars without. Although this
does not preclude that the inner regions of some cold dust systems may be dust free, it disqualifies their host stars
as good targets for an exo-Earth imaging survey. We thus concentrate our analysis on the clean stars in our sample,
meaning stars without cold dust\footnote{We ignore here the presence or absence of hot dust. The origin of this dust is
still unclear and we find no correlation between the presence of hot dust and our detections.}. We reach our best
sensitivity in terms of zodi levels for early type stars, which will dominate our statistics. On the other hand,
Sun-like stars are preferred targets for future exo-Earth imaging missions, because they are more numerous in the solar
neighborhood and stellar suppression requirements become less stringent for detecting Earth-like planets orbiting them
than early type stars. It is unclear if the results for early type stars can be applied to Sun-like stars, and the
similar detection rate for both samples (despite lower sensitivity for Sun-like stars, Sect.~\ref{sect_res_detrate})
suggest they might not. Thus, we perform the statistical analysis of the median zodi level for both the full sample of
clean stars and for clean Sun-like stars only.

\begin{figure*}[t]
 \includegraphics[width=\textwidth]{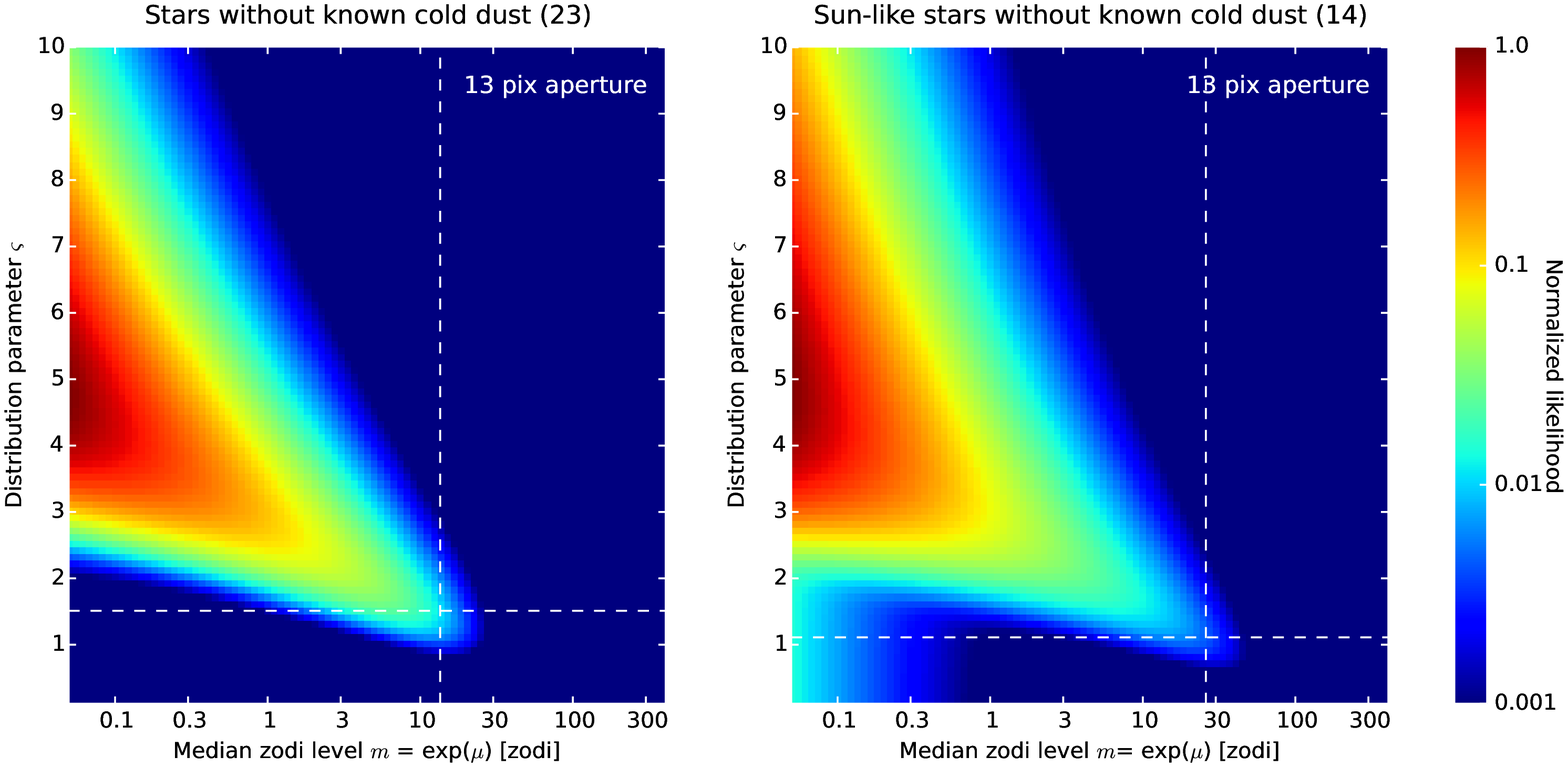}
 \includegraphics[width=\textwidth]{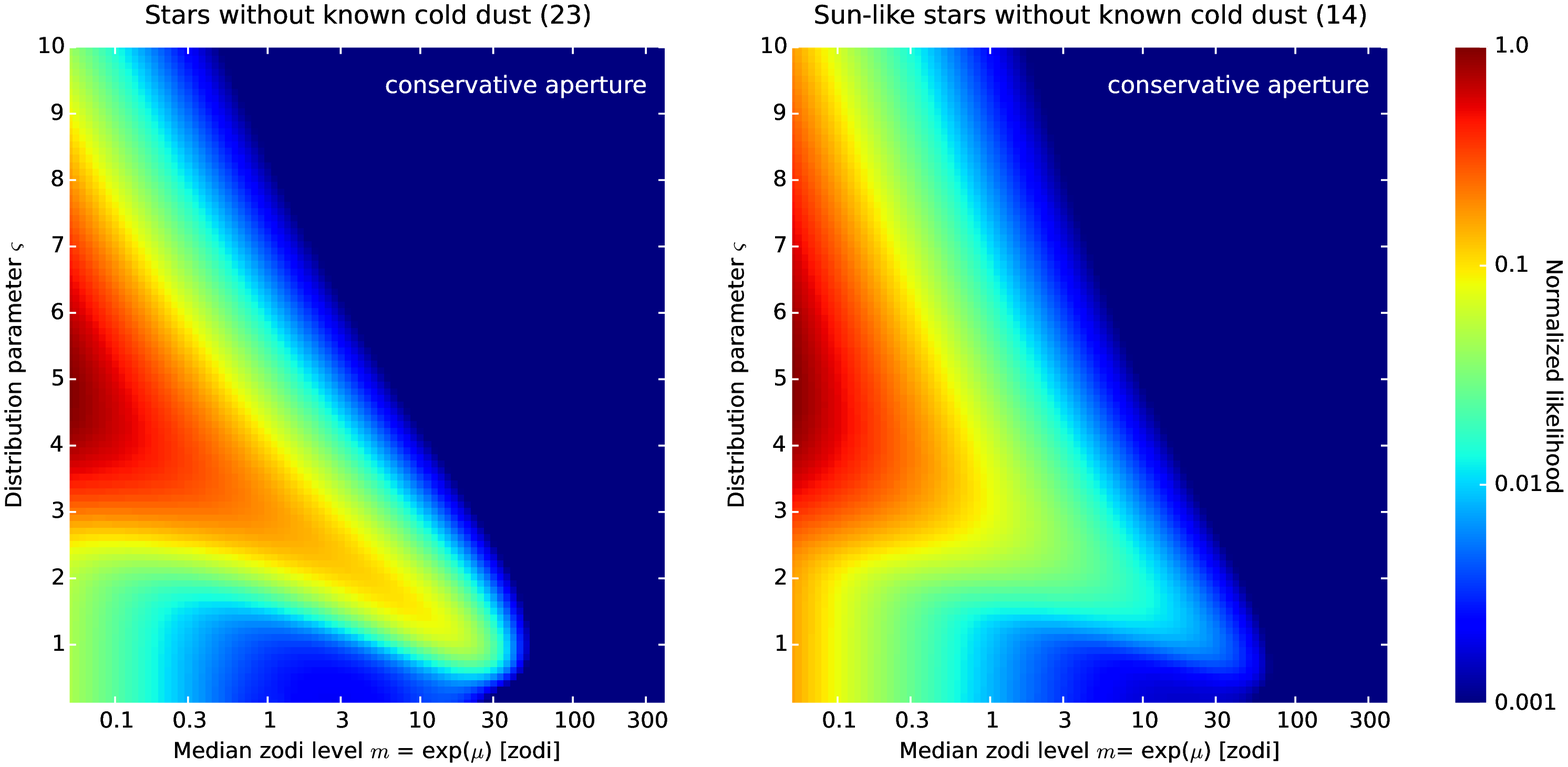}
 \caption{Likelihood distribution of our fits of a lognormal luminosity function to the observed zodi levels and
  uncertainties from Table~\ref{tab_nulls_zodis} for all clean stars (\emph{left}) and all clean Sun-like stars
  (\emph{right}). A $1/m$ prior on the median $m=\exp(\mu)$ is applied, equivalent to assuming a flat prior for the
  lognormal parameter $\mu$. The \emph{top} row shows the results from the zodi levels derived from the 13\,pix
  aperture, while the \emph{bottom} row shows those derived from the conservative aperture. There are no
  inconsistencies between the results derived from the two apertures other than the weaker constraints for the
  conservative aperture due to the larger uncertainties of the individual measurements. The white, dashed lines in the
  plots for the 13\,pix aperture indicate our 95\% confidence interval on $m$ and the corresponding best-fit of the
  sigma parameter of the lognormal luminosity function. These are our current recommendations for the exozodi
  distribution to use for estimating yields for future exo-Earth imaging missions (Sect.~\ref{sect_res_medianzodi}).}
 \label{fig_maps_likelihood}
\end{figure*}

Maps of the likelihood derived for the searched parameter space of $m = \exp\left(\mu\right)$ and $\varsigma$ of the
lognormal luminosity function with the $1/m$ prior applied are shown in Fig.~\ref{fig_maps_likelihood} for both samples
and for the 13\,pix and conservative apertures. Very small values of $\varsigma$ are unable to fit the data well
because our detections of excesses around $\delta$\,UMa and $\theta$\,Boo show that there can still be a significant
amount of warm dust present even for stars without detectable mIR to fIR excess. Thus, a narrow probability
distribution described by a small $\varsigma$ is not able to reproduce the data. A large median is inconsistent with
the large number of non-detections. For intermediate values of $\varsigma$ and $m$, a larger value of $\varsigma$ in
combination with smaller $m$ and vice versa provide fits of similar quality to the data, because both cases are able to
produce a range of excess significances given our sensitivity. We find no inconsistencies between the results for the
13\,pix and conservative apertures, while the constraints from the latter are weaker due to the larger uncertainties on
the individual measurements.

\begin{figure*}[t]
 \includegraphics[width=\linewidth]{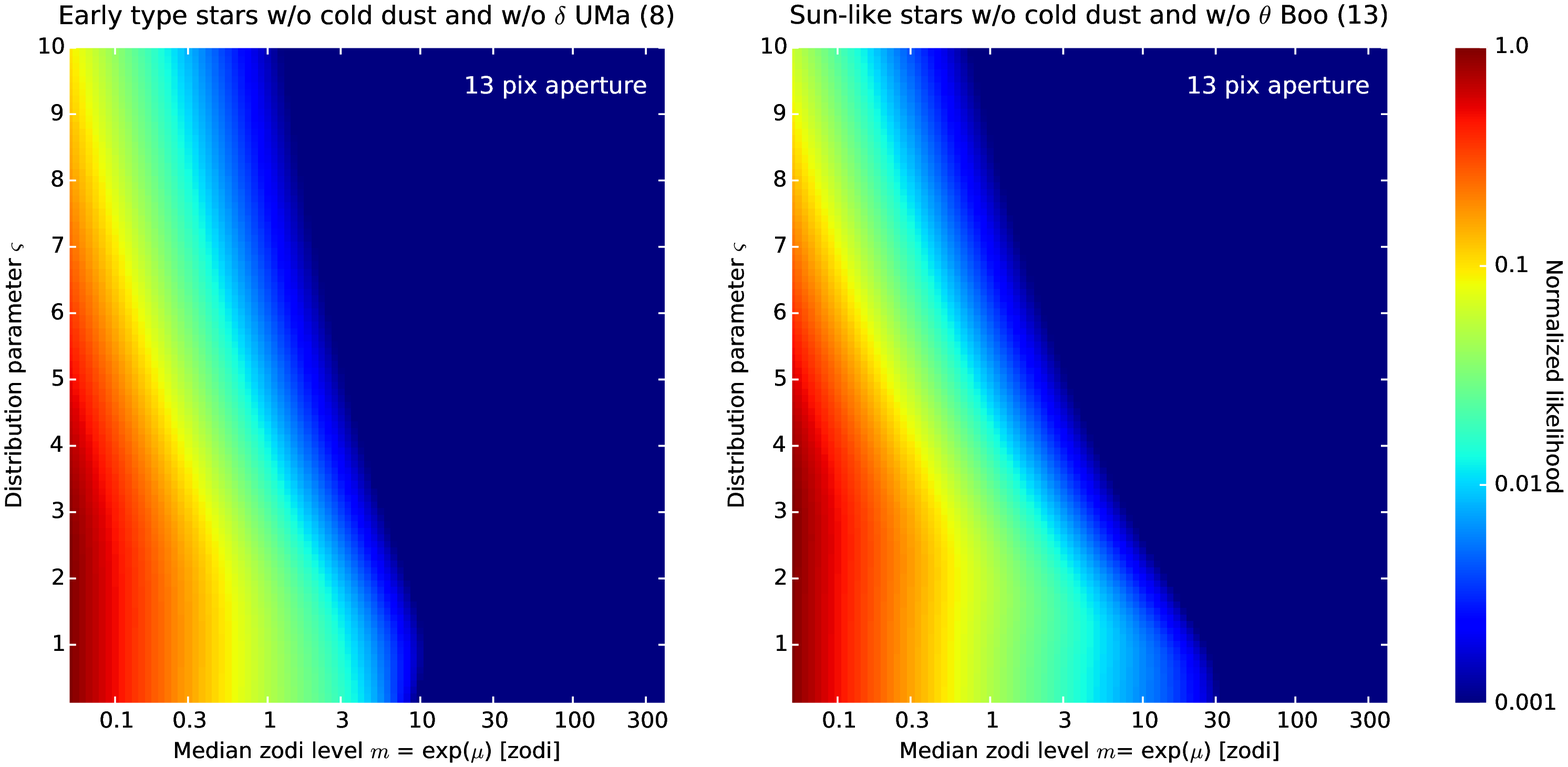}
 \caption{Same as Fig.~\ref{fig_maps_likelihood}, but for the samples of clean early type stars excluding $\delta$\,UMa
  (\emph{left}) and clean Sun-like stars excluding $\theta$\,Boo (\emph{right}).}
 \label{fig_maps_no_exc}
\end{figure*}

The overall shape of our likelihood distribution, and in particular the `nose' of relatively high likelihood around
$\varsigma=2$ and $m=10$ is dominated by the detections around $\delta$\,UMa and $\theta$\,Boo. The detection around
110\,Her would be another such case, if the unclear detection of faint cold dust was considered spurious. Even for
stars without known cold dust we now start to measure the underlying luminosity function. Thus, further increasing the
sensitivity will result in a better measurement of the luminosity function rather than improved upper limits. The shape
of the likelihood distribution also shows a degeneracy in our fits to the data and suggests that a lognormal
distribution might not be a good approximation of the actual luminosity function. We find more evidence for this when
comparing the results from the Bayesian analysis with those from a pure maximum likelihood estimate as performed by
\citet{men14}. While the Bayesian analysis suggests that a zero median is the best fit to the data, the likelihood
peaks at a $m=7^{+8}_{-6}$\,zodis for the full sample of clean stars and $m=13^{+23}_{-12}$\,zodis for clean Sun-like
stars. If there were no degeneracies and the lognormal distribution was a good fit to the data, the results from the
two approaches should be the same. We note, however, that the difference between the two methods is only at the
$\sim $1$\,\sigma$ level and that the two detections around $\delta$\,UMa and $\theta$\,Boo at the $\sim$4$\,\sigma$
level do not allow for any strong conclusions on the actual luminosity function.

\begin{figure*}[t]
 \includegraphics[width=\textwidth]{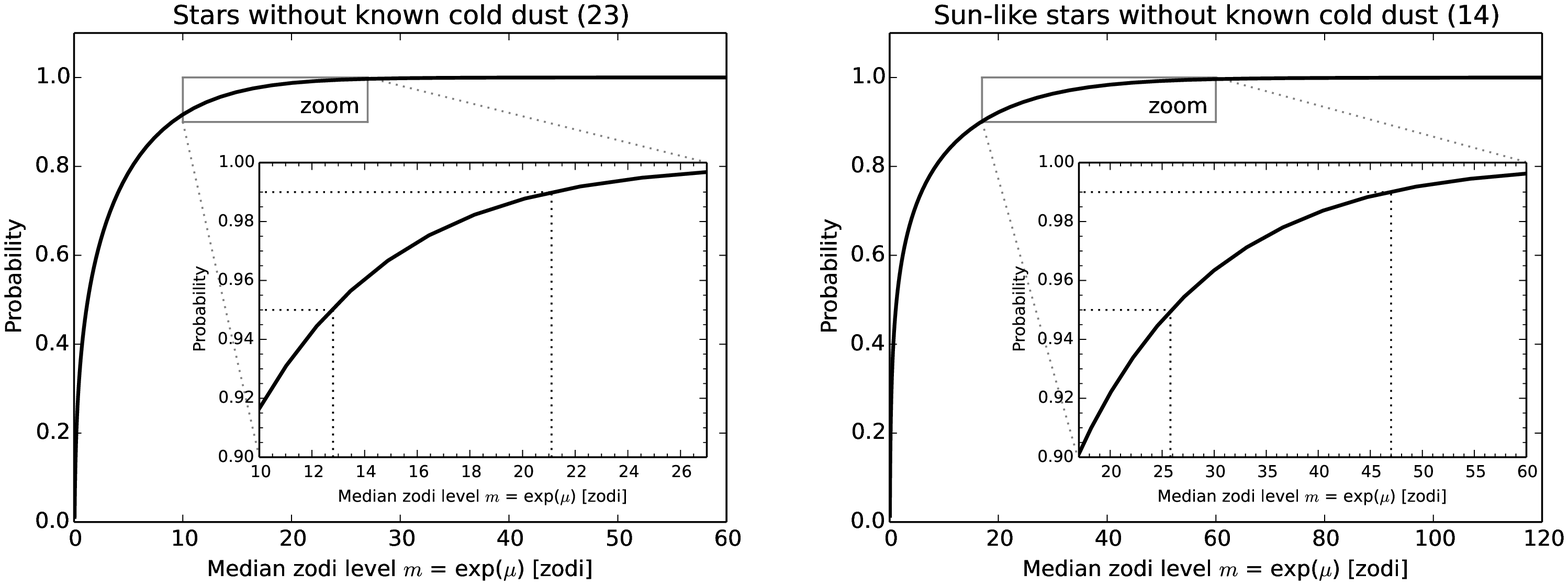}
 \caption{Cumulative probability distribution functions (CPDFs) of the median zodi levels from
  Table~\ref{tab_nulls_zodis} for our sample of all clean stars (\emph{left}) and clean Sun-like stars (\emph{right}).
  The full CPDFs are shown in the large plot and a zoom into the relevant region to determine the 95\% and 99\%
  confidence intervals is shown in the inlay in each plot. Dotted, horizontal and vertical lines mark the 95\% and 99\%
  confidence levels and corresponding median zodi levels.}
 \label{fig_cpdfs}
\end{figure*}

A viable alternative to a broad, lognormal luminosity function or similar single-peaked or monotonous distributions
would be a bimodal one, where the majority of systems have relatively low zodi levels, but a small number of stars are
surrounded by a significant amount of warm dust. Stars without detections of cold dust but with detected HZ dust might,
e.g., harbor a Kuiper belt or asteroid belt analog that is massive enough to sustain a high zodi level through inward
migration of dust grains, but that is too faint to be detected in available data. Stars without large amounts of HZ
dust might not have a cold dust belt at all, or might harbor one or more giant planets between the cold belt and the HZ
that prevent dust from migrating inward in large amounts. Confirming a bimodal luminosity function (e.g., in the light
of potentially higher dust levels around Sun-like stars compared to early type stars) and identifying stars belonging
to the `high zodi level' and `low zodi level' categories would thus be most valuable for our understanding of the
architectures and dynamics of planetary systems. Moreover, it would be favorable for exo-Earth imaging surveys, because
(a) the majority of the targets would have a relatively low zodi level and (b) the stars surrounded by a large amount
of dust could be identified prior to the mission by extensive target vetting with the LBTI and similar instruments
(e.g., the hi-5 concept at the Very Large Telescope Interferometer, \citealt{def18}) in the next two decades.

We illustrate these advantages by excluding $\delta$\,UMa and $\theta$\,Boo from our samples and repeating the median
zodi analysis. We do this for early type and Sun-like stars separately. The resulting likelihood maps are shown in
Fig.~\ref{fig_maps_no_exc}. The nose toward large $m$ and small $\varsigma$ disappears for both samples. The upper
limits on the median zodi levels on these samples thus improve by a factor of $\sim$2. Thus, our upper limits for stars
identified to belong to the `low zodi level' category are approximarely twice as strong as for stars not vetted by LBTI
observations. Discriminating between these two scenarios requires a larger sample and more sensitive observations. Both
can be provided by extending the HOSTS survey (larger sample and better sensitivity due to completed observations and
higher data quality from the setpoint modulation and more experience with the data acquisition compared to some of the
data presented in this work). Specifically, we estimate that completing the observations for all stars in this paper
and observing an equal number of additional stars will suffice to conclude if stars like $\delta$\,UMa and
$\theta$\,Boo are outliers or can be explained by the high excess tail of a lognormal or similar distribution.

From the available data we derive upper limits on the median zodi levels in our samples using the results from our
Bayesian analysis. The CPDFs derived for the the samples of all clean stars and clean Sun-like stars using the 13\,pix
aperture are shown in Fig.~\ref{fig_cpdfs}. For all clean stars we find upper limits of $m=13$\,zodis and $m=21$\,zodis
at 95\% and 99\% confidence. For clean Sun-like stars we find upper limits of $m=26$\,zodis and $m=47$\,zodis,
respectively. From our results, we suggest the use of conservative, but not overly pessimistic assumptions on the
exozodi luminosity function when simulating yields for future exo-Earth imaging surveys. A lognormal distribution with
$m$ equivalent to our 95\% confidence level and the corresponding best-fit value of $\varsigma$ seem appropriate. For
all clean stars, these parameters are $m=13$\,zodi and $\varsigma=1.5$. For clean Sun-like, the parameters are $m=26$
\,zodis and $\varsigma=1.2$.

\subsection{Constraints on the exozodi luminosity function}
\label{sect_lum_funct}

\begin{figure}[t]
 \includegraphics[width=\linewidth]{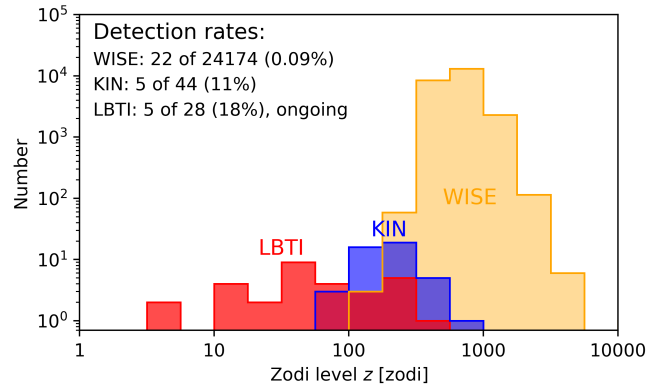}
 \caption{Distribution of sensitivity to HZ dust and sample sizes of LBTI, KIN, and the WISE sample from \citet{ken13}.
  Combining these samples will allow for a comprehensive analysis of the exozodi luminosity function over the range of
  few zodis to its brightest specimen at a several 1,000 zodis.}
 \label{fig_hist_lbti_wise_kin}
\end{figure}

In the previous section we put a constraint on the median zodi level of different samples of stars from our
observations. For this, we assumed a lognormal distribution. Strong constraints on the actual shape of the luminosity
function are not possible based on our limited sample and number of detections. However, our results constrain the
faintest currently reachable regime of the luminosity function and can be combined with available constraints on the
bright end. The cleanest such statistics focussing specifically on HZ dust have been derived from a sample of
Wide-field Infrared Survey Explorer (WISE) observations by \citet{ken13}. We compare the sensitivity to zodi levels,
and sample size of the LBTI, the WISE sample, and the KIN data in Fig.~\ref{fig_hist_lbti_wise_kin}.

\begin{figure}[t]
 \includegraphics[width=\linewidth]{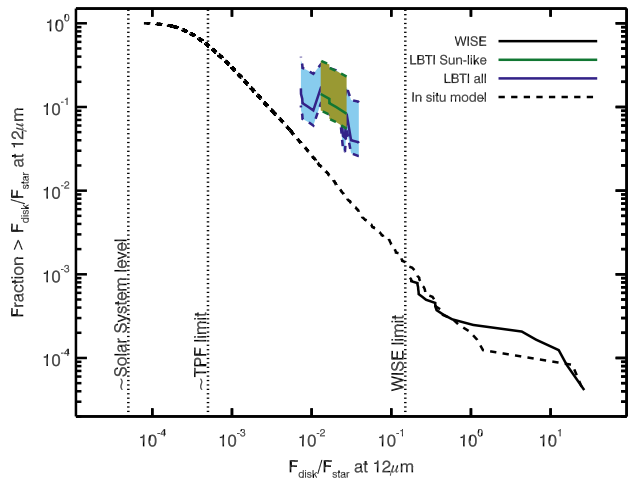}
 \caption{Exozodi luminosity function constrained by our LBTI statistics and the WISE sample by \citet{ken13}.
  Occurrence rates and sensitivities for Sun-like and all stars are consistent, so that the two curved lie on top of
  each other. The slope from the WISE detection rate to the one from the LBTI appears steeper than the prediction from
  the in situ dust production model by \citet{ken13}.}
 \label{fig_lum_func}
\end{figure}

The observable used by \citet{ken13} for the statistical analysis and the presentation of the modeling results is the
disk-to-star flux ratio, rather than the zodi level. We thus first convert our LBTI zodi levels to disk-to-star flux
ratios. We use again our SZ~model, but note that this conversion is more uncertain than deriving the zodi level from
the null measurements, because we now extrapolate from the spatially filtered LBTI excess measurements to photometric
excesses which are more sensitive to the spatial dust distribution (radial slope, inner edge). Converting WISE excesses
to zodi levels as was done to create Fig.~\ref{fig_hist_lbti_wise_kin} is equally affected by the same uncertainties.
Moreover, this would impact already published data, require detailed information on the much larger WISE sample, and
complicate the comparison to the model presented by \citet{ken13}. We then plot the occurrence rate of exozodiacal dust
inferred from our observations for all stars and all Sun-like stars (removing $\eta$\,Crv and $\beta$\,Leo as described
in Sect.~\ref{sect_res_detrate}) over the disk-to-star flux ratio together with the detection rates from \citet{ken13}.
The result is shown in Fig.~\ref{fig_lum_func}. The conversion from zodi level to flux ratio eliminates the sensitivity
advantage to HZ dust of the LBTI for early-type stars.

We also plot the two-component in situ model of an initial dust belt evolving over time and of random, additional dust
production over the star's life time. This model predicts a power-law slope of the luminosity function with an exponent
of -1. We find that our inferred occurrence rate is higher than predicted by the model, a power-law slope of -2 seems
better suited to reproduce the data. This might suggest that an additional dust delivery mechanism is at play for low
dust levels in the range of few zodis to several ten zodis. This could for example be explained by dust delivery
through cometary activity or from an outer dust belt through Poynting-Robertson and stellar wind drag that can only
sustain dust levels up to a maximum surface density for a given system configuration (e.g., \citealt{ken15b}). A
detailed analysis of the samples and evolutionary modeling are necessary for a better understanding of this behavior of
the luminosity function, but are beyond the scope of this paper.

It is important to note that an extrapolation from the WISE and LBTI rates to the occurrence rate at lower zodi levels
cannot be compared to our median zodi analysis in the previous section. For the median zodi analysis we focussed on
stars without known cold dust which are thought to be good targets for exo-Earth imaging, while here we consider all
stars, for which we find a higher detection rate in Sect.~\ref{sect_res_detrate}.

\section{Conclusions}
\label{sect_conc}

We have presented the first statistical results from the HOSTS survey for HZ dust around nearby stars. Our sensitivity
for individual, completed targets is a factor 5~--~10 better than previous observations. Although only a limited sample
of stars have been observed so far, the statistical constraints from our survey are already 2~--~5~times stronger than
earlier results.

We find four new detections, resulting in an overall detection rate of 18\%. This means we are now reaching a
sensitivity at which statistical samples of stars with HZ dust can be created, similar to fIR observations of debris
disks and nIR observations of hot exozodiacal dust. We find the first three detections around Sun-like stars and the
first two around stars without any previously known circumstellar dust. Our inferred occurrence rate at LBTI's
sensitivity is $18^{+9}_{-5}\%$ for all stars and similar for early type stars and Sun-like stars ($17^{+15}_{-6}\%$
vs.\ $19^{+13}_{-6}\%$). It is significantly different for stars with and without a previously known Kuiper belt-like
disk of cold dust ($60^{+16}_{-21}\%$ vs.\ $8^{+10}_{-3}\%$), confirming earlier results at higher sensitivity.
Interestingly, the similar detection rate around early type and Sun-like stars comes at a four times lower sensitivity
to HZ dust around Sun-like stars which might suggest that the HZs of Sun-like stars are in general dustier than those
of early type stars. This tentative result, however, is derived from a small number of detections mostly at the
3~--~5$\,\sigma$ level. It thus needs confirmation from a larger sample and more sensitive observations. A detailed
analysis of the detected systems might also reveal an alternative explanation.

A most puzzling result is our non-detection of warm dust around Vega, for which massive asteroid belt and Kuiper belt
analogs have been detected in the mIR to fIR and a large amount of hot dust has been detected in the nIR. This raises
the question of what mechanism clears the region between $\sim$0.5\,AU and $\sim$5\,AU from the star of dust.

A statistical analysis of our sample -- assuming a lognormal luminosity function -- puts upper limits on the median
zodi level of stars without previously known cold dust to 13\,zodis and 21\,zodis at 95\% and 99\% confidence,
respectively. For Sun-like stars only, the corresponding limits are 26\,zodis and 47\,zodis. We demonstrate that these
limits are no longer governed by the measurement uncertainties as was the case for earlier work, but by the discovery
of a few systems with detected excesses despite the absence of detectable amounts of cold dust. We note the possibility
that the actual, underlying distribution might be bimodal, rather than lognormal, including a few systems with large
amounts of HZ dust and the majority of systems with little such dust. We estimate the limit that can be put on the
median zodi level of a target list for an exo-Earth imaging survey that has been fully vetted by LBTI or similar
observations. We find that it would already be approximately twice as strong as for stars without LBTI observations and
will further improve with more stars being observed. Thus, constraining the exozodi luminosity function by increasing
the available sample size and improving the sensitivity of the observations is critical. Both can be achieved with an
extended HOSTS survey, and extensive vetting of future exo-Earth imaging targets can be done with the LBTI and similar
facilities.

Comparing our inferred occurrence rates of HZ dust in the faint regime of the luminosity function with previous,
photometric results for the bright end suggests that its slope is steeper than predicted by a model of in-situ dust
production. This suggests an additional dust delivery mechanism at lower dust levels than could be detected
photometrically.

From our current results, we suggest to use a lognormal exozodi luminosity function with conservative but not overly
pessimistic parameters when simulating yields for future exo-Earth imaging missions. A combination of our 95\%
confidence upper limit on the median $m$ and the corresponding best-fit value of the sigma parameter $\varsigma$ of the
distribution seem appropriate. From our results using all stars without known cold dust, these parameters are 
$m=13$\,zodi and $\varsigma=1.5$. For Sun-like stars only, the parameters are $m=26$\,zodis and $\varsigma=1.2$.

\acknowledgments

The Large Binocular Telescope Interferometer is funded by the National Aeronautics and Space Administration as part of
its Exoplanet Exploration Program. The LBT is an international collaboration among institutions in the United States,
Italy, and Germany. LBT Corporation partners are: The University of Arizona on behalf of the Arizona university system;
Instituto Nazionale di Astrofisica, Italy; LBT Beteiligungsgesellschaft, Germany, representing the Max-Planck Society,
the Astrophysical Institute Potsdam, and Heidelberg University; The Ohio State University, and The Research
Corporation, on behalf of The University of Notre Dame, University of Minnesota and University of Virginia. This
research has made extensive use of the SIMBAD database \citep{wen00} and the VizieR catalogue access tool
\citep{och00}, both operated at CDS, Strasbourg, France, of Python, including the NumPy, SciPy, Matplotlib
\citep{hun07}, and Astorpy \citep{ast13} libraries, and of NASA's Astrophysics Data System Bibliographic Services. GMK
is supported by the Royal Society as a Royal Society University Research Fellow. AS is partially supported by funding
from the Center for Exoplanets and Habitable Worlds. The Center for Exoplanets and Habitable Worlds is supported by the
Pennsylvania State University, the Eberly College of Science, and the Pennsylvania Space Grant Consortium. JMS is supported
by NASA through Hubble Fellowship grant HST-HF2-51398.001-A awarded by the Space Telescope Science Institute, which is
operated by the Association of Universities for Research in Astronomy, Inc., for NASA, under contract NAS5-26555.

%

\facilities{LBT:LBTI}





\appendix

\startlongtable
\begin{deluxetable}{ccccr@{~\dots~}lc}
\tablecaption{Observing Log\label{tab_obslog}}
\tablewidth{700pt}
\tabletypesize{\small}
\tablehead{ HD \# & Name           & UT Date    & Calibrators HD \#      & \multicolumn{2}{r}{HA range [h]} & Comments} 
\startdata  
10476  & 107\,Psc        & 2016-11-14 & 7087                    & $-0.24$ &  $0.59$  & average data quality \\
       &                 & 2016-11-16 & 7318,   6953            & $-0.71$ &  $0.34$  & mediocre data quality \\
16160  & GJ\,105~A       & 2016-11-15 & 21051,  13596           &  $0.38$ &  $1.04$  & mediocre data quality \\
22049  & $\epsilon$\,Eri & 2014-11-09 & 18322,  29065           &  $2.06$ &  $2.59$  & com., low data quality \\
       &                 & 2014-11-10 & 18322,  23249           &  $0.29$ &  $1.07$  & com., mediocre data quality \\
30652  & 1\,Ori          & 2017-02-09 & 31421,  31767           & $-0.03$ &  $0.99$  & bkg., low data quality \\
33111  & $\beta$\,Eri    & 2017-02-10 & 31767,  36780           &  $1.08$ &  $2.02$  & average data quality \\
34411  & $\lambda$\,Aur  & 2017-01-29 & 38656,  40441           &  $1.55$ &  $3.26$  & high data quality \\
48373  & $\xi$\,Gem      & 2016-11-14 & 49968,  48433           & $-0.09$ &  $0.64$  & average data quality \\
       &                 & 2016-11-15 & 52960                   & $-1.63$ &  $0.70$  & mediocre data quality \\
81937  & 23\,UMa         & 2016-11-15 & 86378                   & $-0.88$ &  $0.17$  & mediocre data quality \\
       &                 & 2017-02-11 & 73108,  92424           & $-2.54$ & $-1.64$  & good data quality\\
88230  & GJ\,380         & 2017-04-06 & 86378,  95212           & $-0.96$ & $-0.27$  & good data quality\\
89449  & 40\,Leo         & 2017-02-09 & 89024,  93257           & $-1.98$ & $-0.36$  & bkg., low data quality \\
95418  & $\beta$\,UMa    & 2017-04-03 & 86378,  94247,  95212   & $-1.16$ &  $0.98$  & good data quality \\
97603  & $\delta$\,Leo   & 2017-02-10 & 99902,  94336           & $-1.61$ & $-0.99$  & good data quality \\
       &                 & 2017-05-12 & 99169,  98262           &  $0.54$ &  $1.22$  & good data quality \\
102647 & $\beta$\,Leo    & 2015-02-08 & 104979, 109742, 108381  &  $1.33$ &  $2.92$  & com., good data quality \\
103287 & $\gamma$\,UMa   & 2017-04-06 & 94247,  95212           & $-0.47$ & $-0.03$  & good data quality \\
       &                 & 2017-05-01 & 102224, 107274          &  $2.09$ &  $2.76$  & good data quality \\ 
106591 & $\delta$\,UMa   & 2017-02-09 & 107465, 102328          & $-0.78$ &  $0.35$  & average data quality \\
       &                 & 2017-05-21 & 101673, 113092          &  $0.60$ &  $1.60$  & mediocre data quality \\
108767 & $\delta$\,Crv   & 2017-02-10 & 114113, 111500          & $-1.19$ & $-0.52$  & mediocre data quality \\
109085 & $\eta$\,Crv     & 2014-02-12 & 108522, 107418, 109272  & $-0.28$ &  $2.18$  & com., low data quality \\
120136 & $\tau$\,Boo     & 2017-05-12 & 114326, 125560          &  $0.62$ &  $2.19$  & average data quality \\
126660 & $\theta$\,Boo   & 2017-02-09 & 128902                  & $-0.29$ &  $0.35$  & bkg., low data quality \\
       &                 & 2017-04-11 & 128902, 138265          & $-0.56$ &  $0.76$  & good data quality \\
128167 & $\sigma$\,Boo   & 2017-04-03 & 133392                  & $-0.39$ &  $0.68$  & low data quality \\
       &                 & 2017-04-06 & 126597, 129972          & $-1.22$ & $-0.59$  & average data quality \\
129502 & $\mu$\,Vir      & 2017-02-10 & 131477, 133165, 130952  & $-1.43$ &  $0.16$  & average data quality \\
141004 & $\lambda$\,Ser  & 2017-05-01 & 145892, 145085          &  $0.24$ &  $0.90$  & good data quality \\
142373 & $\chi$\,Her     & 2017-04-11 & 137704, 144204, 137704  &  $0.83$ &  $2.29$  & good data quality \\
142860 & $\gamma$\,Ser   & 2017-04-06 & 149009, 142574          & $-0.85$ & $-0.18$  & good data quality \\
       &                 & 2017-05-21 & 141992, 145892          & $-0.50$ &  $0.18$  & good data quality \\
172167 & $\alpha$\,Lyr   & 2017-04-06 & 164646, 163770          & $-1.25$ & $-0.61$  & good data quality \\
173667 & 110\,Her        & 2017-04-08 & 170951, 176527          & $-1.69$ & $-0.54$  & average data quality \\
185144 & $\sigma$\,Dra   & 2017-05-01 & 191277, 170693          & $-1.81$ & $-0.84$  & good data quality \\
187642 & $\alpha$\,Aql   & 2017-05-12 & 184406, 189695, 192107  & $-2.65$ &  $0.20$  & sat., bkg., mediocre data
 quality \\
203280 & $\alpha$\,Cep   & 2016-10-16 & 198149, 209960          &  $2.20$ &  $3.11$  & mediocre data quality \\
215648 & $\xi$\,Peg~A    & 2016-11-14 & 218792                  &  $0.47$ &  $1.74$  & mediocre data quality \\
       &                 & 2016-11-16 & 209167, 220009          & $-0.23$ &  $0.73$  & mediocre data quality \\
\enddata                                                                                                     
\tablecomments{Abbreviations in the Comments column are: \emph{com.}~--~commissioning data, \emph{bkg.}~--~data
 affected by strong background variation, \emph{sat.}~--~part of the data unusable due to saturation. In addition to
 these specific cases, the data quality is judged based on the uncertainty of the null measurements.}
\end{deluxetable}




\bibliographystyle{aasjournal}
\bibliography{bibtex}

\begin{thebibliography}{}
\expandafter\ifx\csname natexlab\endcsname\relax\def\natexlab#1{#1}\fi

\bibitem[{{Absil} {et~al.}(2006){Absil}, {di Folco}, {M{\'e}rand}, {Augereau},
  {Coud{\'e} du Foresto}, {Aufdenberg}, {Kervella}, {Ridgway}, {Berger}, {ten
  Brummelaar}, {Sturmann}, {Sturmann}, {Turner}, \& {McAlister}}]{abs06}
{Absil}, O., {di Folco}, E., {M{\'e}rand}, A., {et~al.} 2006, \aap, 452, 237

\bibitem[{{Absil} {et~al.}(2013){Absil}, {Defr{\`e}re}, {Coud{\'e} du Foresto},
  {Di Folco}, {M{\'e}rand}, {Augereau}, {Ertel}, {Hanot}, {Kervella},
  {Mollier}, {Scott}, {Che}, {Monnier}, {Thureau}, {Tuthill}, {ten Brummelaar},
  {McAlister}, {Sturmann}, {Sturmann}, \& {Turner}}]{abs13}
{Absil}, O., {Defr{\`e}re}, D., {Coud{\'e} du Foresto}, V., {et~al.} 2013,
  \aap, 555, A104

\bibitem[{{Akeson} {et~al.}(2009){Akeson}, {Ciardi}, {Millan-Gabet}, {Merand},
  {di Folco}, {Monnier}, {Beichman}, {Absil}, {Aufdenberg}, {McAlister}, {ten
  Brummelaar}, {Sturmann}, {Sturmann}, \& {Turner}}]{ake09}
{Akeson}, R.~L., {Ciardi}, D.~R., {Millan-Gabet}, R., {et~al.} 2009, \apj, 691,
  1896

\bibitem[{{Anglada-Escud{\'e}} \& {Butler}(2012)}]{ang12}
{Anglada-Escud{\'e}}, G., \& {Butler}, R.~P. 2012, \apjs, 200, 15

\bibitem[{{Astropy Collaboration} {et~al.}(2013){Astropy Collaboration},
  {Robitaille}, {Tollerud}, {Greenfield}, {Droettboom}, {Bray}, {Aldcroft},
  {Davis}, {Ginsburg}, {Price-Whelan}, {Kerzendorf}, {Conley}, {Crighton},
  {Barbary}, {Muna}, {Ferguson}, {Grollier}, {Parikh}, {Nair}, {Unther},
  {Deil}, {Woillez}, {Conseil}, {Kramer}, {Turner}, {Singer}, {Fox}, {Weaver},
  {Zabalza}, {Edwards}, {Azalee Bostroem}, {Burke}, {Casey}, {Crawford},
  {Dencheva}, {Ely}, {Jenness}, {Labrie}, {Lim}, {Pierfederici}, {Pontzen},
  {Ptak}, {Refsdal}, {Servillat}, \& {Streicher}}]{ast13}
{Astropy Collaboration}, {Robitaille}, T.~P., {Tollerud}, E.~J., {et~al.} 2013,
  \aap, 558, A33

\bibitem[{{Aufdenberg} {et~al.}(2006){Aufdenberg}, {M{\'e}rand}, {Coud{\'e} du
  Foresto}, {Absil}, {Di Folco}, {Kervella}, {Ridgway}, {Berger}, {ten
  Brummelaar}, {McAlister}, {Sturmann}, {Sturmann}, \& {Turner}}]{auf06}
{Aufdenberg}, J.~P., {M{\'e}rand}, A., {Coud{\'e} du Foresto}, V., {et~al.}
  2006, \apj, 645, 664

\bibitem[{{Aumann}(1985)}]{aum85}
{Aumann}, H.~H. 1985, \pasp, 97, 885

\bibitem[{{Aumann}(1988)}]{aum88}
---. 1988, \aj, 96, 1415

\bibitem[{{Aumann} {et~al.}(1984){Aumann}, {Beichman}, {Gillett}, {de Jong},
  {Houck}, {Low}, {Neugebauer}, {Walker}, \& {Wesselius}}]{aum84}
{Aumann}, H.~H., {Beichman}, C.~A., {Gillett}, F.~C., {et~al.} 1984, \apjl,
  278, L23

\bibitem[{{Backman} {et~al.}(2009){Backman}, {Marengo}, {Stapelfeldt}, {Su},
  {Wilner}, {Dowell}, {Watson}, {Stansberry}, {Rieke}, {Megeath}, {Fazio}, \&
  {Werner}}]{bac09}
{Backman}, D., {Marengo}, M., {Stapelfeldt}, K., {et~al.} 2009, \apj, 690, 1522

\bibitem[{{Backman} \& {Paresce}(1993)}]{bac93}
{Backman}, D.~E., \& {Paresce}, F. 1993, in Protostars and Planets III, ed.
  {E.~H.~Levy \& J.~I.~Lunine}, 1253--1304

\bibitem[{{Bailey} {et~al.}(2014){Bailey}, {Hinz}, {Puglisi}, {Esposito},
  {Vaitheeswaran}, {Skemer}, {Defr{\`e}re}, {Vaz}, \& {Leisenring}}]{bai14}
{Bailey}, V.~P., {Hinz}, P.~M., {Puglisi}, A.~T., {et~al.} 2014, in \procspie,
  Vol. 9148, Adaptive Optics Systems IV, 914803

\bibitem[{{Barnes}(2007)}]{bar07}
{Barnes}, S.~A. 2007, \apj, 669, 1167

\bibitem[{{Beichman} {et~al.}(2006){Beichman}, {Bryden}, {Stapelfeldt},
  {Gautier}, {Grogan}, {Shao}, {Velusamy}, {Lawler}, {Blaylock}, {Rieke},
  {Lunine}, {Fischer}, {Marcy}, {Greaves}, {Wyatt}, {Holland}, \&
  {Dent}}]{bei06}
{Beichman}, C.~A., {Bryden}, G., {Stapelfeldt}, K.~R., {et~al.} 2006, \apj,
  652, 1674

\bibitem[{{Benedict} {et~al.}(2006){Benedict}, {McArthur}, {Gatewood}, {Nelan},
  {Cochran}, {Hatzes}, {Endl}, {Wittenmyer}, {Baliunas}, {Walker}, {Yang},
  {K{\"u}rster}, {Els}, \& {Paulson}}]{ben06}
{Benedict}, G.~F., {McArthur}, B.~E., {Gatewood}, G., {et~al.} 2006, \aj, 132,
  2206

\bibitem[{{B{\"o}hm} {et~al.}(2016){B{\"o}hm}, {Pott}, {Borelli}, {Hinz},
  {Defr{\`e}re}, {Downey}, {Hill}, {Summers}, {Conrad}, {K{\"u}rster},
  {Herbst}, \& {Sawodny}}]{boe16}
{B{\"o}hm}, M., {Pott}, J.-U., {Borelli}, J., {et~al.} 2016, in \procspie, Vol.
  9906, Ground-based and Airborne Telescopes VI, 99062R

\bibitem[{{Bonsor} {et~al.}(2014){Bonsor}, {Raymond}, {Augereau}, \&
  {Ormel}}]{bon14}
{Bonsor}, A., {Raymond}, S.~N., {Augereau}, J.-C., \& {Ormel}, C.~W. 2014,
  \mnras, 441, 2380

\bibitem[{{Booth} {et~al.}(2013){Booth}, {Kennedy}, {Sibthorpe}, {Matthews},
  {Wyatt}, {Duch{\^e}ne}, {Kavelaars}, {Rodriguez}, {Greaves}, {Koning},
  {Vican}, {Rieke}, {Su}, {Moro-Mart{\'{\i}}n}, \& {Kalas}}]{boo13}
{Booth}, M., {Kennedy}, G., {Sibthorpe}, B., {et~al.} 2013, \mnras, 428, 1263

\bibitem[{{Booth} {et~al.}(2017){Booth}, {Dent}, {Jord{\'a}n}, {Lestrade},
  {Hales}, {Wyatt}, {Casassus}, {Ertel}, {Greaves}, {Kennedy}, {Matr{\`a}},
  {Augereau}, \& {Villard}}]{boo17}
{Booth}, M., {Dent}, W.~R.~F., {Jord{\'a}n}, A., {et~al.} 2017, \mnras, 469,
  3200

\bibitem[{{Bord{\'e}} {et~al.}(2002){Bord{\'e}}, {Coud{\'e} du Foresto},
  {Chagnon}, \& {Perrin}}]{bor02}
{Bord{\'e}}, P., {Coud{\'e} du Foresto}, V., {Chagnon}, G., \& {Perrin}, G.
  2002, \aap, 393, 183

\bibitem[{{Brogi} {et~al.}(2009){Brogi}, {Marzari}, \& {Paolicchi}}]{bro09}
{Brogi}, M., {Marzari}, F., \& {Paolicchi}, P. 2009, \aap, 499, L13

\bibitem[{{Chavez-Dagostino} {et~al.}(2016){Chavez-Dagostino}, {Bertone},
  {Cruz-Saenz de Miera}, {Marshall}, {Wilson}, {S{\'a}nchez-Arg{\"u}elles},
  {Hughes}, {Kennedy}, {Vega}, {De la Luz}, {Dent}, {Eiroa}, {G{\'o}mez-Ruiz},
  {Greaves}, {Lizano}, {L{\'o}pez-Valdivia}, {Mamajek}, {Monta{\~n}a},
  {Olmedo}, {Rodr{\'{\i}}guez-Montoya}, {Schloerb}, {Yun}, {Zavala}, \&
  {Zeballos}}]{cha16}
{Chavez-Dagostino}, M., {Bertone}, E., {Cruz-Saenz de Miera}, F., {et~al.}
  2016, \mnras, 462, 2285

\bibitem[{{Chelli} {et~al.}(2016){Chelli}, {Duvert}, {Bourg{\`e}s}, {Mella},
  {Lafrasse}, {Bonneau}, \& {Chesneau}}]{che16}
{Chelli}, A., {Duvert}, G., {Bourg{\`e}s}, L., {et~al.} 2016, \aap, 589, A112

\bibitem[{{Chen} {et~al.}(2006){Chen}, {Sargent}, {Bohac}, {Kim},
  {Leibensperger}, {Jura}, {Najita}, {Forrest}, {Watson}, {Sloan}, \&
  {Keller}}]{che06}
{Chen}, C.~H., {Sargent}, B.~A., {Bohac}, C., {et~al.} 2006, \apjs, 166, 351

\bibitem[{{Churcher} {et~al.}(2011){Churcher}, {Wyatt}, {Duch{\^e}ne},
  {Sibthorpe}, {Kennedy}, {Matthews}, {Kalas}, {Greaves}, {Su}, \&
  {Rieke}}]{chu11}
{Churcher}, L.~J., {Wyatt}, M.~C., {Duch{\^e}ne}, G., {et~al.} 2011, \mnras,
  417, 1715

\bibitem[{{Crooke} {et~al.}(2016){Crooke}, {Roberge}, {Domagal-Goldman},
  {Mandell}, {Bolcar}, {Rioux}, {Perez}, \& {Smith}}]{cro16}
{Crooke}, J.~A., {Roberge}, A., {Domagal-Goldman}, S.~D., {et~al.} 2016, in
  \procspie, Vol. 9904, Space Telescopes and Instrumentation 2016: Optical,
  Infrared, and Millimeter Wave, 99044R

\bibitem[{{Danchi} {et~al.}(2014){Danchi}, {Bailey}, {Bryden}, {Defrere},
  {Haniff}, {Hinz}, {Kennedy}, {Mennesson}, {Millan-Gabet}, {Rieke}, {Roberge},
  {Serabyn}, {Skemer}, {Stapelfeldt}, {Weinberger}, \& {Wyatt}}]{dan14}
{Danchi}, W., {Bailey}, V., {Bryden}, G., {et~al.} 2014, in \procspie, Vol.
  9146, Optical and Infrared Interferometry IV, 914607

\bibitem[{{Defr{\`e}re} {et~al.}(2010){Defr{\`e}re}, {Absil}, {den Hartog},
  {Hanot}, \& {Stark}}]{def10}
{Defr{\`e}re}, D., {Absil}, O., {den Hartog}, R., {Hanot}, C., \& {Stark}, C.
  2010, \aap, 509, A9

\bibitem[{{Defr{\`e}re} {et~al.}(2011){Defr{\`e}re}, {Absil}, {Augereau}, {di
  Folco}, {Berger}, {Coud{\'e} du Foresto}, {Kervella}, {Le Bouquin},
  {Lebreton}, {Millan-Gabet}, {Monnier}, {Olofsson}, \& {Traub}}]{def11}
{Defr{\`e}re}, D., {Absil}, O., {Augereau}, J.-C., {et~al.} 2011, \aap, 534, A5

\bibitem[{{Defr{\`e}re} {et~al.}(2012){Defr{\`e}re}, {Lebreton}, {Le Bouquin},
  {Lagrange}, {Absil}, {Augereau}, {Berger}, {di Folco}, {Ertel}, {Kluska},
  {Montagnier}, {Millan-Gabet}, {Traub}, \& {Zins}}]{def12}
{Defr{\`e}re}, D., {Lebreton}, J., {Le Bouquin}, J.-B., {et~al.} 2012, \aap,
  546, L9

\bibitem[{{Defr{\`e}re} {et~al.}(2015){Defr{\`e}re}, {Hinz}, {Skemer},
  {Kennedy}, {Bailey}, {Hoffmann}, {Mennesson}, {Millan-Gabet}, {Danchi},
  {Absil}, {Arbo}, {Beichman}, {Brusa}, {Bryden}, {Downey}, {Durney},
  {Esposito}, {Gaspar}, {Grenz}, {Haniff}, {Hill}, {Lebreton}, {Leisenring},
  {Males}, {Marion}, {McMahon}, {Montoya}, {Morzinski}, {Pinna}, {Puglisi},
  {Rieke}, {Roberge}, {Serabyn}, {Sosa}, {Stapeldfeldt}, {Su}, {Vaitheeswaran},
  {Vaz}, {Weinberger}, \& {Wyatt}}]{def15}
{Defr{\`e}re}, D., {Hinz}, P.~M., {Skemer}, A.~J., {et~al.} 2015, \apj, 799, 42

\bibitem[{{Defr{\`e}re} {et~al.}(2016){Defr{\`e}re}, {Hinz}, {Mennesson},
  {Hoffmann}, {Millan-Gabet}, {Skemer}, {Bailey}, {Danchi}, {Downey}, {Durney},
  {Grenz}, {Hill}, {McMahon}, {Montoya}, {Spalding}, {Vaz}, {Absil}, {Arbo},
  {Bailey}, {Brusa}, {Bryden}, {Esposito}, {Gaspar}, {Haniff}, {Kennedy},
  {Leisenring}, {Marion}, {Nowak}, {Pinna}, {Powell}, {Puglisi}, {Rieke},
  {Roberge}, {Serabyn}, {Sosa}, {Stapeldfeldt}, {Su}, {Weinberger}, \&
  {Wyatt}}]{def16}
{Defr{\`e}re}, D., {Hinz}, P.~M., {Mennesson}, B., {et~al.} 2016, \apj, 824, 66

\bibitem[{{Defr{\`e}re} {et~al.}(2018){Defr{\`e}re}, {Absil}, {Berger},
  {Boulet}, {Danchi}, {Ertel}, {Gallenne}, {H{\'e}nault}, {Hinz}, {Huby},
  {Ireland}, {Kraus}, {Labadie}, {Le Bouquin}, {Martin}, {Matter},
  {M{\'e}rand}, {Mennesson}, {Minardi}, {Monnier}, {Norris}, {Orban de Xivry},
  {Pedretti}, {Pott}, {Reggiani}, {Serabyn}, {Surdej}, {Tristram}, \&
  {Woillez}}]{def18}
{Defr{\`e}re}, D., {Absil}, O., {Berger}, J.-P., {et~al.} 2018, ArXiv e-prints,
  arXiv:1801.04148

\bibitem[{{Dermott} {et~al.}(2002){Dermott}, {Kehoe}, {Durda}, {Grogan}, \&
  {Nesvorn{\'y}}}]{der02}
{Dermott}, S.~F., {Kehoe}, T.~J.~J., {Durda}, D.~D., {Grogan}, K., \&
  {Nesvorn{\'y}}, D. 2002, in ESA Special Publication, Vol. 500, Asteroids,
  Comets, and Meteors: ACM 2002, ed. B.~{Warmbein}, 319--322

\bibitem[{{Duch{\^e}ne} {et~al.}(2014){Duch{\^e}ne}, {Arriaga}, {Wyatt},
  {Kennedy}, {Sibthorpe}, {Lisse}, {Holland}, {Wisniewski}, {Clampin}, {Kalas},
  {Pinte}, {Wilner}, {Booth}, {Horner}, {Matthews}, \& {Greaves}}]{duc14}
{Duch{\^e}ne}, G., {Arriaga}, P., {Wyatt}, M., {et~al.} 2014, \apj, 784, 148

\bibitem[{{Durkan} {et~al.}(2016){Durkan}, {Janson}, \& {Carson}}]{dur16}
{Durkan}, S., {Janson}, M., \& {Carson}, J.~C. 2016, \apj, 824, 58

\bibitem[{{Eiroa} {et~al.}(2011){Eiroa}, {Marshall}, {Mora}, {Krivov},
  {Montesinos}, {Absil}, {Ardila}, {Ar{\'e}valo}, {Augereau}, {Bayo}, {Danchi},
  {Del Burgo}, {Ertel}, {Fridlund}, {Gonz{\'a}lez-Garc{\'{\i}}a}, {Heras},
  {Lebreton}, {Liseau}, {Maldonado}, {Meeus}, {Montes}, {Pilbratt}, {Roberge},
  {Sanz-Forcada}, {Stapelfeldt}, {Th{\'e}bault}, {White}, \& {Wolf}}]{eir11}
{Eiroa}, C., {Marshall}, J.~P., {Mora}, A., {et~al.} 2011, \aap, 536, L4

\bibitem[{{Eiroa} {et~al.}(2013){Eiroa}, {Marshall}, {Mora}, {Montesinos},
  {Absil}, {Augereau}, {Bayo}, {Bryden}, {Danchi}, {del Burgo}, {Ertel},
  {Fridlund}, {Heras}, {Krivov}, {Launhardt}, {Liseau}, {L{\"o}hne},
  {Maldonado}, {Pilbratt}, {Roberge}, {Rodmann}, {Sanz-Forcada}, {Solano},
  {Stapelfeldt}, {Th{\'e}bault}, {Wolf}, {Ardila}, {Ar{\'e}valo}, {Beichmann},
  {Faramaz}, {Gonz{\'a}lez-Garc{\'{\i}}a}, {Guti{\'e}rrez}, {Lebreton},
  {Mart{\'{\i}}nez-Arn{\'a}iz}, {Meeus}, {Montes}, {Olofsson}, {Su}, {White},
  {Barrado}, {Fukagawa}, {Gr{\"u}n}, {Kamp}, {Lorente}, {Morbidelli},
  {M{\"u}ller}, {Mutschke}, {Nakagawa}, {Ribas}, \& {Walker}}]{eir13}
---. 2013, \aap, 555, A11

\bibitem[{{Ertel} {et~al.}(2012){Ertel}, {Wolf}, \& {Rodmann}}]{ert12b}
{Ertel}, S., {Wolf}, S., \& {Rodmann}, J. 2012, \aap, 544, A61

\bibitem[{{Ertel} {et~al.}(2014){Ertel}, {Absil}, {Defr{\`e}re}, {Le Bouquin},
  {Augereau}, {Marion}, {Blind}, {Bonsor}, {Bryden}, {Lebreton}, \&
  {Milli}}]{ert14b}
{Ertel}, S., {Absil}, O., {Defr{\`e}re}, D., {et~al.} 2014, \aap, 570, A128

\bibitem[{{Ertel} {et~al.}(2016){Ertel}, {Defr{\`e}re}, {Absil}, {Le Bouquin},
  {Augereau}, {Berger}, {Blind}, {Bonsor}, {Lagrange}, {Lebreton}, {Marion},
  {Milli}, \& {Olofsson}}]{ert16}
{Ertel}, S., {Defr{\`e}re}, D., {Absil}, O., {et~al.} 2016, \aap, 595, A44

\bibitem[{{Faramaz} {et~al.}(2017){Faramaz}, {Ertel}, {Booth}, {Cuadra}, \&
  {Simmonds}}]{far17}
{Faramaz}, V., {Ertel}, S., {Booth}, M., {Cuadra}, J., \& {Simmonds}, C. 2017,
  \mnras, 465, 2352

\bibitem[{{Fischer} {et~al.}(2014){Fischer}, {Marcy}, \& {Spronck}}]{fis14}
{Fischer}, D.~A., {Marcy}, G.~W., \& {Spronck}, J.~F.~P. 2014, \apjs, 210, 5

\bibitem[{{Frolov}(1970)}]{fro70}
{Frolov}, M.~S. 1970, Information Bulletin on Variable Stars, 427

\bibitem[{{G{\'a}sp{\'a}r} \& {Rieke}(2014)}]{gas14}
{G{\'a}sp{\'a}r}, A., \& {Rieke}, G.~H. 2014, \apj, 784, 33

\bibitem[{{G{\'a}sp{\'a}r} {et~al.}(2013){G{\'a}sp{\'a}r}, {Rieke}, \&
  {Balog}}]{gas13}
{G{\'a}sp{\'a}r}, A., {Rieke}, G.~H., \& {Balog}, Z. 2013, \apj, 768, 25

\bibitem[{{Gezari} {et~al.}(1993){Gezari}, {Schmitz}, {Pitts}, \&
  {Mead}}]{gez93}
{Gezari}, D.~Y., {Schmitz}, M., {Pitts}, P.~S., \& {Mead}, J.~M. 1993, {Catalog
  of infrared observations, third edition}

\bibitem[{{Gillett}(1986)}]{gil86}
{Gillett}, F.~C. 1986, in Astrophysics and Space Science Library, Vol. 124,
  Light on Dark Matter, ed. F.~P. {Israel}, 61--69

\bibitem[{{Greaves} {et~al.}(1998){Greaves}, {Holland}, {Moriarty-Schieven},
  {Jenness}, {Dent}, {Zuckerman}, {McCarthy}, {Webb}, {Butner}, {Gear}, \&
  {Walker}}]{gre98}
{Greaves}, J.~S., {Holland}, W.~S., {Moriarty-Schieven}, G., {et~al.} 1998,
  \apjl, 506, L133

\bibitem[{{Greaves} {et~al.}(2014){Greaves}, {Sibthorpe}, {Acke}, {Pantin},
  {Vandenbussche}, {Olofsson}, {Dominik}, {Barlow}, {Bendo}, {Blommaert},
  {Brandeker}, {de Vries}, {Dent}, {Di Francesco}, {Fridlund}, {Gear},
  {Harvey}, {Hogerheijde}, {Holland}, {Ivison}, {Liseau}, {Matthews},
  {Pilbratt}, {Walker}, \& {Waelkens}}]{gre14}
{Greaves}, J.~S., {Sibthorpe}, B., {Acke}, B., {et~al.} 2014, \apjl, 791, L11

\bibitem[{{Gulliver} {et~al.}(1994){Gulliver}, {Hill}, \& {Adelman}}]{gul94}
{Gulliver}, A.~F., {Hill}, G., \& {Adelman}, S.~J. 1994, \apjl, 429, L81

\bibitem[{{Hanot} {et~al.}(2011){Hanot}, {Mennesson}, {Martin}, {Liewer},
  {Loya}, {Mawet}, {Riaud}, {Absil}, \& {Serabyn}}]{han11}
{Hanot}, C., {Mennesson}, B., {Martin}, S., {et~al.} 2011, \apj, 729, 110

\bibitem[{{Hatzes} {et~al.}(2000){Hatzes}, {Cochran}, {McArthur}, {Baliunas},
  {Walker}, {Campbell}, {Irwin}, {Yang}, {K{\"u}rster}, {Endl}, {Els},
  {Butler}, \& {Marcy}}]{hat00}
{Hatzes}, A.~P., {Cochran}, W.~D., {McArthur}, B., {et~al.} 2000, \apjl, 544,
  L145

\bibitem[{{Hinz} {et~al.}(2016){Hinz}, {Defr{\`e}re}, {Skemer}, {Bailey},
  {Stone}, {Spalding}, {Vaz}, {Pinna}, {Puglisi}, {Esposito}, {Montoya},
  {Downey}, {Leisenring}, {Durney}, {Hoffmann}, {Hill}, {Millan-Gabet},
  {Mennesson}, {Danchi}, {Morzinski}, {Grenz}, {Skrutskie}, \& {Ertel}}]{hin16}
{Hinz}, P.~M., {Defr{\`e}re}, D., {Skemer}, A., {et~al.} 2016, in \procspie,
  Vol. 9907, Optical and Infrared Interferometry and Imaging V, 990704

\bibitem[{{Hoffmann} {et~al.}(2014){Hoffmann}, {Hinz}, {Defr{\`e}re},
  {Leisenring}, {Skemer}, {Arbo}, {Montoya}, \& {Mennesson}}]{hof14}
{Hoffmann}, W.~F., {Hinz}, P.~M., {Defr{\`e}re}, D., {et~al.} 2014, in
  \procspie, Vol. 9147, Ground-based and Airborne Instrumentation for Astronomy
  V, 91471O

\bibitem[{{Howard} \& {Fulton}(2016)}]{how16}
{Howard}, A.~W., \& {Fulton}, B.~J. 2016, \pasp, 128, 114401

\bibitem[{Hunter(2007)}]{hun07}
Hunter, J.~D. 2007, Computing In Science \& Engineering, 9, 90

\bibitem[{{Ibukiyama} \& {Arimoto}(2002)}]{ibu02}
{Ibukiyama}, A., \& {Arimoto}, N. 2002, \aap, 394, 927

\bibitem[{{Jones} {et~al.}(2015){Jones}, {White}, {Boyajian}, {Schaefer},
  {Baines}, {Ireland}, {Patience}, {ten Brummelaar}, {McAlister}, {Ridgway},
  {Sturmann}, {Sturmann}, {Turner}, {Farrington}, \& {Goldfinger}}]{jon16}
{Jones}, J., {White}, R.~J., {Boyajian}, T., {et~al.} 2015, \apj, 813, 58

\bibitem[{{Kelsall} {et~al.}(1998){Kelsall}, {Weiland}, {Franz}, {Reach},
  {Arendt}, {Dwek}, {Freudenreich}, {Hauser}, {Moseley}, {Odegard},
  {Silverberg}, \& {Wright}}]{kel98}
{Kelsall}, T., {Weiland}, J.~L., {Franz}, B.~A., {et~al.} 1998, \apj, 508, 44

\bibitem[{{Kennedy} \& {Piette}(2015)}]{ken15b}
{Kennedy}, G.~M., \& {Piette}, A. 2015, \mnras, 449, 2304

\bibitem[{{Kennedy} \& {Wyatt}(2013)}]{ken13}
{Kennedy}, G.~M., \& {Wyatt}, M.~C. 2013, \mnras, 433, 2334

\bibitem[{{Kennedy} {et~al.}(2015){Kennedy}, {Wyatt}, {Bailey}, {Bryden},
  {Danchi}, {Defr{\`e}re}, {Haniff}, {Hinz}, {Lebreton}, {Mennesson},
  {Millan-Gabet}, {Morales}, {Pani{\'c}}, {Rieke}, {Roberge}, {Serabyn},
  {Shannon}, {Skemer}, {Stapelfeldt}, {Su}, \& {Weinberger}}]{ken15}
{Kennedy}, G.~M., {Wyatt}, M.~C., {Bailey}, V., {et~al.} 2015, \apjs, 216, 23

\bibitem[{{Kharchenko} {et~al.}(2007){Kharchenko}, {Scholz}, {Piskunov},
  {Roeser}, \& {Schilbach}}]{kha07}
{Kharchenko}, N.~V., {Scholz}, R., {Piskunov}, A.~E., {Roeser}, S., \&
  {Schilbach}, E. 2007, VizieR Online Data Catalog, 3254, 0

\bibitem[{{Kirchschlager} {et~al.}(2017){Kirchschlager}, {Wolf}, {Krivov},
  {Mutschke}, \& {Brunngr{\"a}ber}}]{kir17}
{Kirchschlager}, F., {Wolf}, S., {Krivov}, A.~V., {Mutschke}, H., \&
  {Brunngr{\"a}ber}, R. 2017, \mnras, 467, 1614

\bibitem[{{Koerner} {et~al.}(2010){Koerner}, {Kim}, {Trilling}, {Larson},
  {Cotera}, {Stapelfeldt}, {Wahhaj}, {Fajardo-Acosta}, {Padgett}, \&
  {Backman}}]{koe10}
{Koerner}, D.~W., {Kim}, S., {Trilling}, D.~E., {et~al.} 2010, \apjl, 710, L26

\bibitem[{{Krist} {et~al.}(2016){Krist}, {Nemati}, \& {Mennesson}}]{kri16}
{Krist}, J., {Nemati}, B., \& {Mennesson}, B. 2016, Journal of Astronomical
  Telescopes, Instruments, and Systems, 2, 011003

\bibitem[{{Krivov} {et~al.}(2013){Krivov}, {Eiroa}, {L{\"o}hne}, {Marshall},
  {Montesinos}, {del Burgo}, {Absil}, {Ardila}, {Augereau}, {Bayo}, {Bryden},
  {Danchi}, {Ertel}, {Lebreton}, {Liseau}, {Mora}, {Mustill}, {Mutschke},
  {Neuh{\"a}user}, {Pilbratt}, {Roberge}, {Schmidt}, {Stapelfeldt},
  {Th{\'e}bault}, {Vitense}, {White}, \& {Wolf}}]{kri13}
{Krivov}, A.~V., {Eiroa}, C., {L{\"o}hne}, T., {et~al.} 2013, \apj, 772, 32

\bibitem[{{Lawler} {et~al.}(2009){Lawler}, {Beichman}, {Bryden}, {Ciardi},
  {Tanner}, {Su}, {Stapelfeldt}, {Lisse}, \& {Harker}}]{law09}
{Lawler}, S.~M., {Beichman}, C.~A., {Bryden}, G., {et~al.} 2009, \apj, 705, 89

\bibitem[{{Lebreton} {et~al.}(2016){Lebreton}, {Beichman}, {Bryden},
  {Defr{\`e}re}, {Mennesson}, {Millan-Gabet}, \& {Boccaletti}}]{leb16}
{Lebreton}, J., {Beichman}, C., {Bryden}, G., {et~al.} 2016, \apj, 817, 165

\bibitem[{{Lebreton} {et~al.}(2013){Lebreton}, {van Lieshout}, {Augereau},
  {Absil}, {Mennesson}, {Kama}, {Dominik}, {Bonsor}, {Vandeportal}, {Beust},
  {Defr{\`e}re}, {Ertel}, {Faramaz}, {Hinz}, {Kral}, {Lagrange}, {Liu}, \&
  {Th{\'e}bault}}]{leb13}
{Lebreton}, J., {van Lieshout}, R., {Augereau}, J.-C., {et~al.} 2013, \aap,
  555, A146

\bibitem[{{Lestrade} \& {Thilliez}(2015)}]{les15}
{Lestrade}, J.-F., \& {Thilliez}, E. 2015, \aap, 576, A72

\bibitem[{{Lisse} {et~al.}(2012){Lisse}, {Wyatt}, {Chen}, {Morlok}, {Watson},
  {Manoj}, {Sheehan}, {Currie}, {Thebault}, \& {Sitko}}]{lis12}
{Lisse}, C.~M., {Wyatt}, M.~C., {Chen}, C.~H., {et~al.} 2012, \apj, 747, 93

\bibitem[{{Liu} {et~al.}(2004){Liu}, {Hinz}, {Hoffmann}, {Brusa}, {Wildi},
  {Miller}, {Lloyd-Hart}, {Kenworthy}, {McGuire}, \& {Angel}}]{liu04c}
{Liu}, W.~M., {Hinz}, P.~M., {Hoffmann}, W.~F., {et~al.} 2004, \apjl, 610, L125

\bibitem[{{MacGregor} {et~al.}(2015){MacGregor}, {Wilner}, {Andrews},
  {Lestrade}, \& {Maddison}}]{macgre15}
{MacGregor}, M.~A., {Wilner}, D.~J., {Andrews}, S.~M., {Lestrade}, J.-F., \&
  {Maddison}, S. 2015, \apj, 809, 47

\bibitem[{{Mallik} {et~al.}(2003){Mallik}, {Parthasarathy}, \& {Pati}}]{mal03}
{Mallik}, S.~V., {Parthasarathy}, M., \& {Pati}, A.~K. 2003, \aap, 409, 251

\bibitem[{{Mamajek} \& {Hillenbrand}(2008)}]{mam08}
{Mamajek}, E.~E., \& {Hillenbrand}, L.~A. 2008, \apj, 687, 1264

\bibitem[{{Marino} {et~al.}(2017){Marino}, {Wyatt}, {Pani{\'c}}, {Matr{\`a}},
  {Kennedy}, {Bonsor}, {Kral}, {Dent}, {Duchene}, {Wilner}, {Lisse},
  {Lestrade}, \& {Matthews}}]{mar17b}
{Marino}, S., {Wyatt}, M.~C., {Pani{\'c}}, O., {et~al.} 2017, \mnras, 465, 2595

\bibitem[{{Marion} {et~al.}(2018){Marion}, {Absil}, {Ertel}, {Defr{\`e}re}, {Le
  Bouquin}, Milli, {Blind}, \& {Augereau}}]{mar17}
{Marion}, L., {Absil}, O., {Ertel}, S., {et~al.} 2018, submitted to \aap

\bibitem[{{Marshall} {et~al.}(2013){Marshall}, {Krivov}, {del Burgo}, {Eiroa},
  {Mora}, {Montesinos}, {Ertel}, {Bryden}, {Liseau}, {Augereau}, {Bayo},
  {Danchi}, {L{\"o}hne}, {Maldonado}, {Pilbratt}, {Stapelfeldt}, {Thebault},
  {White}, \& {Wolf}}]{mar13}
{Marshall}, J.~P., {Krivov}, A.~V., {del Burgo}, C., {et~al.} 2013, \aap, 557,
  A58

\bibitem[{{Matthews} {et~al.}(2010){Matthews}, {Sibthorpe}, {Kennedy},
  {Phillips}, {Churcher}, {Duch{\^e}ne}, {Greaves}, {Lestrade}, {Moro-Martin},
  {Wyatt}, {Bastien}, {Biggs}, {Bouvier}, {Butner}, {Dent}, {di Francesco},
  {Eisl{\"o}ffel}, {Graham}, {Harvey}, {Hauschildt}, {Holland}, {Horner},
  {Ibar}, {Ivison}, {Johnstone}, {Kalas}, {Kavelaars}, {Rodriguez}, {Udry},
  {van der Werf}, {Wilner}, \& {Zuckerman}}]{mat10}
{Matthews}, B.~C., {Sibthorpe}, B., {Kennedy}, G., {et~al.} 2010, \aap, 518,
  L135+

\bibitem[{{Mennesson} {et~al.}(2011){Mennesson}, {Serabyn}, {Hanot}, {Martin},
  {Liewer}, \& {Mawet}}]{men11}
{Mennesson}, B., {Serabyn}, E., {Hanot}, C., {et~al.} 2011, \apj, 736, 14

\bibitem[{{Mennesson} {et~al.}(2013){Mennesson}, {Absil}, {Lebreton},
  {Augereau}, {Serabyn}, {Colavita}, {Millan-Gabet}, {Liu}, {Hinz}, \&
  {Th{\'e}bault}}]{men13}
{Mennesson}, B., {Absil}, O., {Lebreton}, J., {et~al.} 2013, \apj, 763, 119

\bibitem[{{Mennesson} {et~al.}(2014){Mennesson}, {Millan-Gabet}, {Serabyn},
  {Colavita}, {Absil}, {Bryden}, {Wyatt}, {Danchi}, {Defr{\`e}re}, {Dor{\'e}},
  {Hinz}, {Kuchner}, {Ragland}, {Scott}, {Stapelfeldt}, {Traub}, \&
  {Woillez}}]{men14}
{Mennesson}, B., {Millan-Gabet}, R., {Serabyn}, E., {et~al.} 2014, \apj, 797,
  119

\bibitem[{{Mennesson} {et~al.}(2016{\natexlab{a}}){Mennesson}, {Defr{\`e}re},
  {Nowak}, {Hinz}, {Millan-Gabet}, {Absil}, {Bailey}, {Bryden}, {Danchi},
  {Kennedy}, {Marion}, {Roberge}, {Serabyn}, {Skemer}, {Stapelfeldt},
  {Weinberger}, \& {Wyatt}}]{men16a}
{Mennesson}, B., {Defr{\`e}re}, D., {Nowak}, M., {et~al.} 2016{\natexlab{a}},
  in \procspie, Vol. 9907, Optical and Infrared Interferometry and Imaging V,
  99070X

\bibitem[{{Mennesson} {et~al.}(2016{\natexlab{b}}){Mennesson}, {Gaudi},
  {Seager}, {Cahoy}, {Domagal-Goldman}, {Feinberg}, {Guyon}, {Kasdin},
  {Marois}, {Mawet}, {Tamura}, {Mouillet}, {Prusti}, {Quirrenbach}, {Robinson},
  {Rogers}, {Scowen}, {Somerville}, {Stapelfeldt}, {Stern}, {Still},
  {Turnbull}, {Booth}, {Kiessling}, {Kuan}, \& {Warfield}}]{men16b}
{Mennesson}, B., {Gaudi}, S., {Seager}, S., {et~al.} 2016{\natexlab{b}}, in
  \procspie, Vol. 9904, Space Telescopes and Instrumentation 2016: Optical,
  Infrared, and Millimeter Wave, 99040L

\bibitem[{{M{\'e}rand} {et~al.}(2005){M{\'e}rand}, {Bord{\'e}}, \& {Coud{\'e}
  du Foresto}}]{mer05}
{M{\'e}rand}, A., {Bord{\'e}}, P., \& {Coud{\'e} du Foresto}, V. 2005, \aap,
  433, 1155

\bibitem[{{Meshkat} {et~al.}(2015){Meshkat}, {Kenworthy}, {Reggiani}, {Quanz},
  {Mamajek}, \& {Meyer}}]{mes15}
{Meshkat}, T., {Kenworthy}, M.~A., {Reggiani}, M., {et~al.} 2015, \mnras, 453,
  2533

\bibitem[{{Millan-Gabet} {et~al.}(2011){Millan-Gabet}, {Serabyn}, {Mennesson},
  {Stark}, {Ragland}, {Hrynevych}, {Woillez}, {Stapelfeldt}, {Bryden},
  {Colavita}, \& {Booth}}]{mil11}
{Millan-Gabet}, R., {Serabyn}, E., {Mennesson}, B., {et~al.} 2011, \apj, 734,
  67

\bibitem[{{Moerchen} {et~al.}(2010){Moerchen}, {Telesco}, \& {Packham}}]{moe10}
{Moerchen}, M.~M., {Telesco}, C.~M., \& {Packham}, C. 2010, \apj, 723, 1418

\bibitem[{{Montesinos} {et~al.}(2016){Montesinos}, {Eiroa}, {Krivov},
  {Marshall}, {Pilbratt}, {Liseau}, {Mora}, {Maldonado}, {Wolf}, {Ertel},
  {Bayo}, {Augereau}, {Heras}, {Fridlund}, {Danchi}, {Solano}, {Kirchschlager},
  {del Burgo}, \& {Montes}}]{mon16}
{Montesinos}, B., {Eiroa}, C., {Krivov}, A.~V., {et~al.} 2016, \aap, 593, A51

\bibitem[{{Moran} {et~al.}(2004){Moran}, {Kuchner}, \& {Holman}}]{mor04}
{Moran}, S.~M., {Kuchner}, M.~J., \& {Holman}, M.~J. 2004, \apj, 612, 1163

\bibitem[{{Nesvorn{\'y}} {et~al.}(2010){Nesvorn{\'y}}, {Jenniskens}, {Levison},
  {Bottke}, {Vokrouhlick{\'y}}, \& {Gounelle}}]{nes10}
{Nesvorn{\'y}}, D., {Jenniskens}, P., {Levison}, H.~F., {et~al.} 2010, \apj,
  713, 816

\bibitem[{{Nu\~nez} {et~al.}(2017){Nu\~nez}, {Scott}, {Mennesson}, {Absil},
  {Augereau}, {Bryden}, {ten Brummelaar}, {Ertel}, {Coude du Foresto},
  {Ridgway}, {Sturmann}, {Sturmann}, {Turner}, \& {Turner}}]{nun17}
{Nu\~nez}, P.~D., {Scott}, N.~J., {Mennesson}, B., {et~al.} 2017, ArXiv
  e-prints, arXiv:1709.01655

\bibitem[{{Ochsenbein} {et~al.}(2000){Ochsenbein}, {Bauer}, \&
  {Marcout}}]{och00}
{Ochsenbein}, F., {Bauer}, P., \& {Marcout}, J. 2000, \aaps, 143, 23

\bibitem[{{Peterson} {et~al.}(2006){Peterson}, {Hummel}, {Pauls}, {Armstrong},
  {Benson}, {Gilbreath}, {Hindsley}, {Hutter}, {Johnston}, {Mozurkewich}, \&
  {Schmitt}}]{pet06}
{Peterson}, D.~M., {Hummel}, C.~A., {Pauls}, T.~A., {et~al.} 2006, \nat, 440,
  896

\bibitem[{{Reidemeister} {et~al.}(2011){Reidemeister}, {Krivov}, {Stark},
  {Augereau}, {L{\"o}hne}, \& {M{\"u}ller}}]{rei11}
{Reidemeister}, M., {Krivov}, A.~V., {Stark}, C.~C., {et~al.} 2011, \aap, 527,
  A57+

\bibitem[{{Rieke} {et~al.}(2016){Rieke}, {G{\'a}sp{\'a}r}, \&
  {Ballering}}]{rie16}
{Rieke}, G.~H., {G{\'a}sp{\'a}r}, A., \& {Ballering}, N.~P. 2016, \apj, 816, 50

\bibitem[{{Rieke} {et~al.}(2005){Rieke}, {Su}, {Stansberry}, {Trilling},
  {Bryden}, {Muzerolle}, {White}, {Gorlova}, {Young}, {Beichman},
  {Stapelfeldt}, \& {Hines}}]{rie05}
{Rieke}, G.~H., {Su}, K.~Y.~L., {Stansberry}, J.~A., {et~al.} 2005, \apj, 620,
  1010

\bibitem[{{Roberge} {et~al.}(2012){Roberge}, {Chen}, {Millan-Gabet},
  {Weinberger}, {Hinz}, {Stapelfeldt}, {Absil}, {Kuchner}, \& {Bryden}}]{rob12}
{Roberge}, A., {Chen}, C.~H., {Millan-Gabet}, R., {et~al.} 2012, \pasp, 124,
  799

\bibitem[{{Serabyn} {et~al.}(2012){Serabyn}, {Mennesson}, {Colavita},
  {Koresko}, \& {Kuchner}}]{ser12}
{Serabyn}, E., {Mennesson}, B., {Colavita}, M.~M., {Koresko}, C., \& {Kuchner},
  M.~J. 2012, \apj, 748, 55

\bibitem[{{Smith} {et~al.}(2009){Smith}, {Wyatt}, \& {Haniff}}]{smi09}
{Smith}, R., {Wyatt}, M.~C., \& {Haniff}, C.~A. 2009, \aap, 503, 265

\bibitem[{{Stark} \& {Kuchner}(2008)}]{sta08}
{Stark}, C.~C., \& {Kuchner}, M.~J. 2008, \apj, 686, 637

\bibitem[{{Stark} \& {Kuchner}(2009)}]{sta09b}
---. 2009, \apj, 707, 543

\bibitem[{{Stark} {et~al.}(2015){Stark}, {Roberge}, {Mandell}, {Clampin},
  {Domagal-Goldman}, {McElwain}, \& {Stapelfeldt}}]{sta15}
{Stark}, C.~C., {Roberge}, A., {Mandell}, A., {et~al.} 2015, \apj, 808, 149

\bibitem[{{Stark} {et~al.}(2016){Stark}, {Shaklan}, {Lisman}, {Cady},
  {Savransky}, {Roberge}, \& {Mandell}}]{sta16}
{Stark}, C.~C., {Shaklan}, S., {Lisman}, D., {et~al.} 2016, Journal of
  Astronomical Telescopes, Instruments, and Systems, 2, 041204

\bibitem[{{Stock} {et~al.}(2010){Stock}, {Su}, {Liu}, {Hinz}, {Rieke},
  {Marengo}, {Stapelfeldt}, {Hines}, \& {Trilling}}]{sto10}
{Stock}, N.~D., {Su}, K.~Y.~L., {Liu}, W., {et~al.} 2010, \apj, 724, 1238

\bibitem[{{Su} {et~al.}(2006){Su}, {Rieke}, {Stansberry}, {Bryden},
  {Stapelfeldt}, {Trilling}, {Muzerolle}, {Beichman}, {Moro-Martin}, {Hines},
  \& {Werner}}]{su06}
{Su}, K.~Y.~L., {Rieke}, G.~H., {Stansberry}, J.~A., {et~al.} 2006, \apj, 653,
  675

\bibitem[{{Su} {et~al.}(2013){Su}, {Rieke}, {Malhotra}, {Stapelfeldt},
  {Hughes}, {Bonsor}, {Wilner}, {Balog}, {Watson}, {Werner}, \&
  {Misselt}}]{su13}
{Su}, K.~Y.~L., {Rieke}, G.~H., {Malhotra}, R., {et~al.} 2013, \apj, 763, 118

\bibitem[{{Su} {et~al.}(2017){Su}, {De Buizer}, {Rieke}, {Krivov}, {L{\"o}hne},
  {Marengo}, {Stapelfeldt}, {Ballering}, \& {Vacca}}]{su17}
{Su}, K.~Y.~L., {De Buizer}, J.~M., {Rieke}, G.~H., {et~al.} 2017, \aj, 153,
  226

\bibitem[{{Thureau} {et~al.}(2014){Thureau}, {Greaves}, {Matthews}, {Kennedy},
  {Phillips}, {Booth}, {Duch{\^e}ne}, {Horner}, {Rodriguez}, {Sibthorpe}, \&
  {Wyatt}}]{thu14}
{Thureau}, N.~D., {Greaves}, J.~S., {Matthews}, B.~C., {et~al.} 2014, \mnras,
  445, 2558

\bibitem[{{Trilling} {et~al.}(2008){Trilling}, {Bryden}, {Beichman}, {Rieke},
  {Su}, {Stansberry}, {Blaylock}, {Stapelfeldt}, {Beeman}, \& {Haller}}]{tri08}
{Trilling}, D.~E., {Bryden}, G., {Beichman}, C.~A., {et~al.} 2008, \apj, 674,
  1086

\bibitem[{{van Leeuwen}(2007)}]{vanle07}
{van Leeuwen}, F. 2007, \aap, 474, 653

\bibitem[{{van Lieshout} {et~al.}(2014){van Lieshout}, {Dominik}, {Kama}, \&
  {Min}}]{vanlie14}
{van Lieshout}, R., {Dominik}, C., {Kama}, M., \& {Min}, M. 2014, ArXiv
  e-prints, arXiv:1404.3271

\bibitem[{{Vican}(2012)}]{vic12}
{Vican}, L. 2012, \aj, 143, 135

\bibitem[{{Weinberger} {et~al.}(2015){Weinberger}, {Bryden}, {Kennedy},
  {Roberge}, {Defr{\`e}re}, {Hinz}, {Millan-Gabet}, {Rieke}, {Bailey},
  {Danchi}, {Haniff}, {Mennesson}, {Serabyn}, {Skemer}, {Stapelfeldt}, \&
  {Wyatt}}]{wei15}
{Weinberger}, A.~J., {Bryden}, G., {Kennedy}, G.~M., {et~al.} 2015, \apjs, 216,
  24

\bibitem[{{Wenger} {et~al.}(2000){Wenger}, {Ochsenbein}, {Egret}, {Dubois},
  {Bonnarel}, {Borde}, {Genova}, {Jasniewicz}, {Lalo{\"e}}, {Lesteven}, \&
  {Monier}}]{wen00}
{Wenger}, M., {Ochsenbein}, F., {Egret}, D., {et~al.} 2000, \aaps, 143, 9

\bibitem[{{Wyatt}(2005)}]{wya05}
{Wyatt}, M.~C. 2005, \aap, 433, 1007

\bibitem[{{Wyatt} {et~al.}(2005){Wyatt}, {Greaves}, {Dent}, \&
  {Coulson}}]{wya05b}
{Wyatt}, M.~C., {Greaves}, J.~S., {Dent}, W.~R.~F., \& {Coulson}, I.~M. 2005,
  \apj, 620, 492

\bibitem[{{Wyatt} {et~al.}(2007){Wyatt}, {Smith}, {Greaves}, {Beichman},
  {Bryden}, \& {Lisse}}]{wya07b}
{Wyatt}, M.~C., {Smith}, R., {Greaves}, J.~S., {et~al.} 2007, \apj, 658, 569

\bibitem[{{Yoon} {et~al.}(2010){Yoon}, {Peterson}, {Kurucz}, \&
  {Zagarello}}]{yoo10}
{Yoon}, J., {Peterson}, D.~M., {Kurucz}, R.~L., \& {Zagarello}, R.~J. 2010,
  \apj, 708, 71

\bibitem[{{Yoon} {et~al.}(2008){Yoon}, {Peterson}, {Zagarello}, {Armstrong}, \&
  {Pauls}}]{yoo08}
{Yoon}, J., {Peterson}, D.~M., {Zagarello}, R.~J., {Armstrong}, J.~T., \&
  {Pauls}, T. 2008, \apj, 681, 570

\bibitem[{{Zuckerman} {et~al.}(2011){Zuckerman}, {Rhee}, {Song}, \&
  {Bessell}}]{zuc11}
{Zuckerman}, B., {Rhee}, J.~H., {Song}, I., \& {Bessell}, M.~S. 2011, \apj,
  732, 61

\end{thebibliography}





\end{document}